\documentclass[journal]{IEEEtran}
\ifCLASSINFOpdf
  \usepackage[pdftex]{graphicx}
  \graphicspath{{../pdf/}{../jpeg/}}
\else
\fi
\usepackage{amsmath}
\usepackage{amsfonts}
\usepackage{array}
\ifCLASSOPTIONcompsoc
  \usepackage[caption=false,font=normalsize,labelfont=sf,textfont=sf]{subfig}
\else
  \usepackage[caption=false,font=footnotesize]{subfig}
\fi

\usepackage{graphicx, booktabs, multirow,xfrac}
\usepackage[inline]{enumitem}
\usepackage{stfloats}

\usepackage{hyperref}
\newcommand{\etal}{\textit{et al.}}

\usepackage{siunitx}

\usepackage{cleveref}

\begin{document}
%
% paper title
% Titles are generally capitalized except for words such as a, an, and, as,
% at, but, by, for, in, nor, of, on, or, the, to and up, which are usually
% not capitalized unless they are the first or last word of the title.
% Linebreaks \\ can be used within to get better formatting as desired.
% Do not put math or special symbols in the title.
\title{Towards duration robust weakly supervised sound event detection}
%
%
% author names and IEEE memberships
% note positions of commas and nonbreaking spaces ( ~ ) LaTeX will not break
% a structure at a ~ so this keeps an author's name from being broken across
% two lines.
% use \thanks{} to gain access to the first footnote area
% a separate \thanks must be used for each paragraph as LaTeX2e's \thanks
% was not built to handle multiple paragraphs
%

\author{Heinrich Dinkel,~\IEEEmembership{Student Member,~IEEE,}
        Mengyue~Wu,~\IEEEmembership{Member,~IEEE,}
        and~Kai~Yu,~\IEEEmembership{Senior Member,~IEEE,}
        % <-this % stops a space
        }
% \thanks{M. Shell was with the Department
% of Electrical and Computer Engineering, Georgia Institute of Technology, Atlanta,
% GA, 30332 USA e-mail: (see http://www.michaelshell.org/contact.html).}% <-this % stops a space
% \thanks{J. Doe and J. Doe are with Anonymous University.}% <-this % stops a space
% \thanks{Manuscript received April 19, 2005; revised August 26, 2015.}}

% note the % following the last \IEEEmembership and also \thanks - 
% these prevent an unwanted space from occurring between the last author name
% and the end of the author line. i.e., if you had this:
% 
% \author{....lastname \thanks{...} \thanks{...} }
%                     ^------------^------------^----Do not want these spaces!
%
% a space would be appended to the last name and could cause every name on that
% line to be shifted left slightly. This is one of those "LaTeX things". For
% instance, "\textbf{A} \textbf{B}" will typeset as "A B" not "AB". To get
% "AB" then you have to do: "\textbf{A}\textbf{B}"
% \thanks is no different in this regard, so shield the last } of each \thanks
% that ends a line with a % and do not let a space in before the next \thanks.
% Spaces after \IEEEmembership other than the last one are OK (and needed) as
% you are supposed to have spaces between the names. For what it is worth,
% this is a minor point as most people would not even notice if the said evil
% space somehow managed to creep in.

% The paper headers
\markboth{Journal of \LaTeX\ Class Files,~Vol.~14, No.~8, August~2015}%
{Shell \MakeLowercase{\textit{et al.}}: Bare Demo of IEEEtran.cls for IEEE Journals}
% The only time the second header will appear is for the odd numbered pages
% after the title page when using the twoside option.
% 
% *** Note that you probably will NOT want to include the author's ***
% *** name in the headers of peer review papers.                   ***
% You can use \ifCLASSOPTIONpeerreview for conditional compilation here if
% you desire.

% If you want to put a publisher's ID mark on the page you can do it like
% thi
%\IEEEpubid{0000--0000/00\$00.00~\copyright~2015 IEEE}
% Remember, if you use this you must call \IEEEpubidadjcol in the second
% column for its text to clear the IEEEpubid mark.

% use for special paper notices
%\IEEEspecialpapernotice{(Invited Paper)}

% make the title area
\maketitle

% As a general rule, do not put math, special symbols or citations
% in the abstract or keywords.
\begin{abstract}
Sound event detection (SED) is the task of tagging the absence or presence of audio events and their corresponding interval within a given audio clip.
While SED can be done using supervised machine learning, where training data is fully labeled with access to per event timestamps and duration, our work focuses on weakly-supervised sound event detection (WSSED), where prior knowledge about an event's duration is unavailable.
Recent research within the field focuses on improving segment- and event-level localization performance for specific datasets regarding specific evaluation metrics.
Specifically, well-performing event-level localization requires fully labeled development subsets to obtain event duration estimates, which significantly benefits localization performance.
Moreover, well-performing segment-level localization models output predictions at a coarse-scale (e.g., 1 second), hindering their deployment on datasets containing very short events ($<$ 1 second).
This work proposes a duration robust CRNN (CDur) framework, which aims to achieve competitive performance in terms of segment- and event-level localization. 
This paper proposes a new post-processing strategy named ``Triple Threshold'' and investigates two data augmentation methods along with a label smoothing method within the scope of WSSED. 
Evaluation of our model is done on the DCASE2017 and 2018 Task 4 datasets, and URBAN-SED.
Our model outperforms other approaches on the DCASE2018 and URBAN-SED datasets without requiring prior duration knowledge.
In particular, our model is capable of similar performance to strongly-labeled supervised models on the URBAN-SED dataset.
Lastly, ablation experiments to reveal that without post-processing, our model's localization performance drop is significantly lower compared with other approaches.
\end{abstract}

% Note that keywords are not normally used for peerreview papers.

\IEEEpeerreviewmaketitle

\section{Introduction}
% Here we have the typical use of a "T" for an initial drop letter
% and "HIS" in caps to complete the first word.
\IEEEPARstart{S}{ound} event detection (SED) research classifies and localizes particular audio events (e.g., dog barking, alarm ringing) within an audio clip, assigning each event a label along with a start point (onset) and an endpoint (offset). 
Label assignment is usually referred to as tagging, while the onset/offset detection is referred to as localization. 
SED can be used for query-based sound retrieval~\cite{Font2018}, smart cities, and homes~\cite{Bello2018, Krstulovic2018}, as well as voice activity detection~\cite{Dinkel2020}. 
Unlike common classification tasks such as image or speaker recognition, a single audio clip might contain multiple different sound events (multi-output), sometimes occurring simultaneously (multi-label). 
In particular, the localization task escalates the difficulty within the scope of SED, since 
different sound events have various time lengths, and each occurrence is unique.
Two main approaches exist to train an effective localization model: Fully supervised SED and weakly supervised SED (WSSED).
Fully supervised approaches, which potentially perform better than weakly supervised ones, require manual time-stamp labeling.
However, manual labeling is a significant hindrance for scaling to large datasets due to the expensive labor cost.
This paper primarily focuses on WSSED, which only has access to clip event labels during training yet requires to predict onsets and offsets at the inference stage.

Challenges such as the Detection and Classification of Acoustic Scenes and Events (DCASE) exemplify the difficulties in training robust SED systems.
DCASE challenge datasets are real-world recordings (e.g., audio with no quality control and lossy compression), thus containing unknown noises and scenarios.
Specifically, in each challenge since 2017, at least one task was primarily concerned with WSSED.
Most previous work focuses on providing single target task-specific solutions for WSSED on either \mbox{tagging-,} \mbox{segment-} or event-level.
Tagging-level solutions are often capable of localizing event boundaries, yet their temporal consistency is subpar to segment- and event-level methods.
This has been seen during the DCASE2017 challenge, where no single model could win both tagging and localization subtasks. 
Solutions optimized for segment level often utilize a fixed target time resolution (e.g., 1 Hz), inhibiting fine-scale localization performance  (e.g., 50 Hz).
Lastly, successful event-level solutions require prior knowledge about each events' duration to obtain temporally consistent predictions.
Previous work in~\cite{Dinkel2019} showed that successful models such as the DCASE2018 task 4 winner are biased towards predicting tags from long-duration clips, which might limit themselves from generalizing towards different datasets (e.g., deploy the same model on a new dataset) since new datasets possibly contain short or unknown duration events.  
In contrast, we aim to enhance WSSED performance, specifically in duration estimation regarding short, abrupt events, without a pre-estimation of each respective event's individual weight.
%Therefore, current state-of-the-art models are still a far 

\section{Related Work}

Most current approaches within SED and WSSED utilize neural networks, in particular convolutional neural networks~\cite{lin2019specialized,McFee2018} (CNN) and convolutional recurrent neural networks~\cite{Dinkel2019,Dinkel2020} (CRNN).
CNN models generally excel at audio tagging~\cite{Kong2019b,Kong2019a} and scale with data, yet falling behind CRNN approaches in onset and offset estimations~\cite{Kothinti2018}.

\begin{figure}
    \centering
    \captionsetup{justification=centering}
    \includegraphics[width=\linewidth]{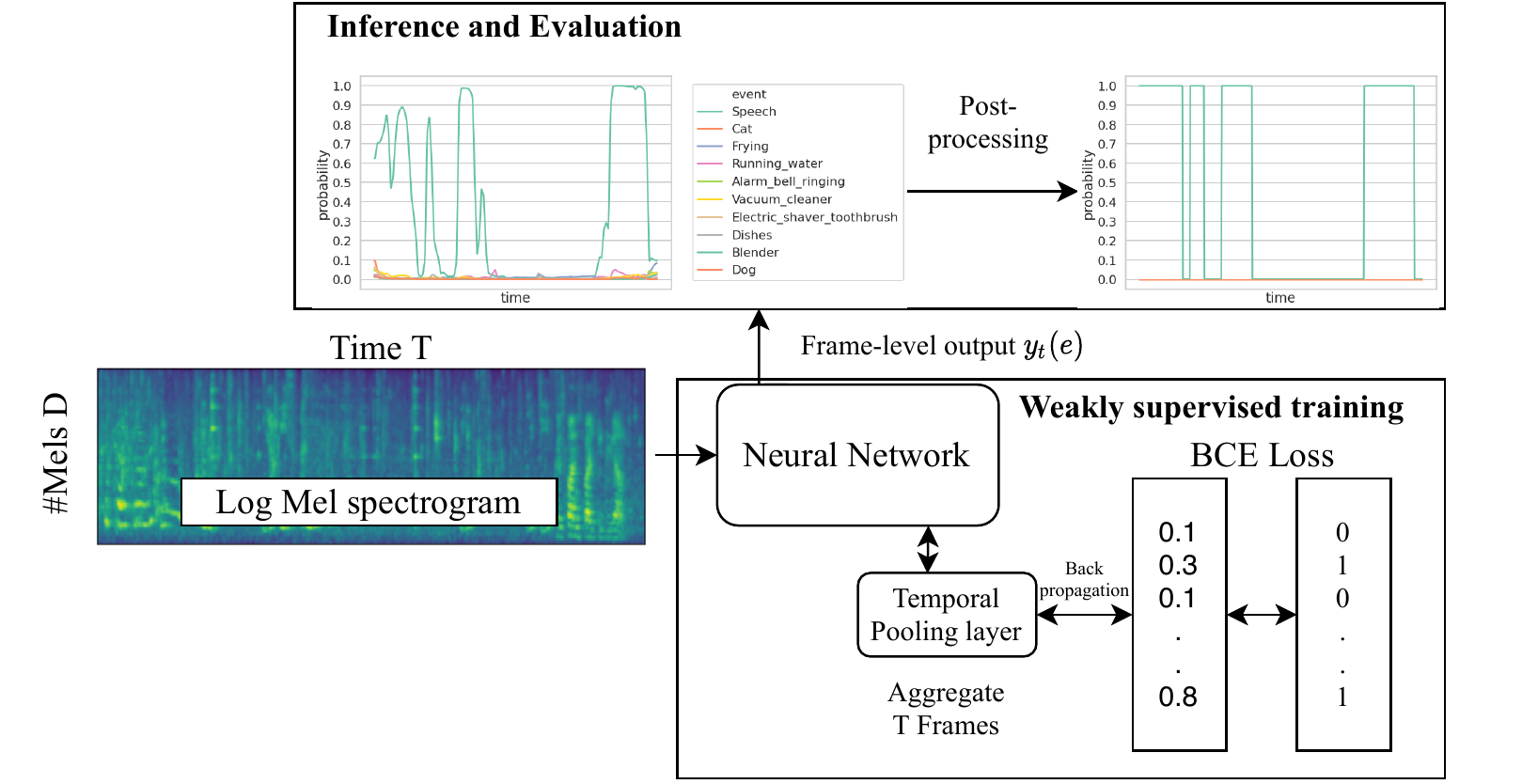}
    \caption{Basic WSSED framework used in our work. During training only the aggregated per event probabilities are learned via LinSoft, while during inference additional per event time-stamps need to be estimated and enhanced via post-processing.}
    \label{fig:framework}
\end{figure}

Apart from different modeling methods, many recent works propose other approaches for the localization conundrum. 
A plethora of temporal pooling strategies are proposed, aiming to summarize frame-level beliefs into a single clip-wise probability. 
McFee \etal~\cite{McFee2018} analyzes the common problem of temporal mean and max pooling for WSSED, indicating that temporal max-pooling preserves event boundaries well; however, it can only backpropagate a single weight. 
Therefore,~\cite{McFee2018} proposed soft-max pooling, an interpolation between mean and max pooling, as well as AutoPool, which is a per-event weighted soft-max pooling approach.
Subsequent work from Wang \etal~\cite{Wang2018} extends the idea of~\cite{McFee2018} and compares other similar pooling methods such as attention to two newly proposed approaches, namely linear and exponential softmax.
Their results on the DCASE2017 task4 dataset indicate that linear softmax could be potentially superior to attention and max-pooling methods in terms of false positives.
By contrast, recent work in~\cite{Kao2020} reports that temporal max-pooling seems to be advantageous in rare sound event scenarios. 
However, though different pooling strategies have been proposed, the correlation between the pooling methods and the classifier backends might be under investigation. 
Attention level pooling methods seem to be preferred for models without sequential model capabilities~\cite{lin2019specialized,Huang2020} (e.g., CNN), whereas max and linear softmax pooling functions have seen success in CRNN frameworks~\cite{Wang2018,Dinkel2019}.

In addition to pooling strategies, a recent popular approach in dealing with different duration lengths is disentanglement.
Lin \etal~\cite{lin2019specialized} presents impressive results on the DCASE2018 task 4 and DCASE19 task 4 datasets, achieving 38.6\% and 42.7\% Event-F1, respectively, by estimating an individual weight sequence for each individual sound event.
Moreover, it is indicated that feature-level aggregation methods (e.g., hidden layer) should be preferred over event-level methods (e.g., output layer). 
Recent work in~\cite{Huang2020} proposes multi-branch learning, similar to multi-task learning, which utilizes a multitude of temporal pooling strategies in order to prevent overfitting towards a single method. 
Similar methods regarding disentanglement are spectral event-specific masking, as introduced in~\cite{Kong2018}.
The main idea is to estimate time-frequency masks via a CNN regarding a target sound event.
The advantage of this method is that it can jointly localize as well as filter audio clips.

Lastly, another research category within the field focuses on constructing frame-level losses during training.
An example of such a method is~\cite{Pellegrini2019}, which introduces a cosine penalty between different time-event predictions, aiming to enhance the per time step discriminability of each event.
This idea is similar to large margin softmax (L-softmax)~\cite{liu2016large} and has resulted in an F1 score of 32.42\% on the DCASE2018 task4 development dataset.

The success of approaches such as~\cite{lin2019specialized,Lu2018,Lee2017b} relies on prior knowledge about event-wise duration or a target resolution. These methods, in turn, have the following downsides: 1) They are costly since manual labor is required for time-stamp labeling and 2) they rely on consistent duration labeling for each event between available development and unseen evaluation sets, as well as 3) become near impossible to estimate for large quantities of event labels and lastly 4) can only be used for their specific purpose, e.g., tagging, event-level or 1 s segment estimation.
Besides, previously introduced models~\cite{lin2019specialized,Lu2018,Lee2017b}, which excel on a specific dataset, have yet to be shown to work on others successfully.

\paragraph*{Contribution}

In our work, we modify and extend the framework of~\cite{Dinkel2019} further towards other datasets and aim to analyze the benefits and the limits of duration robust training.
Our main goal with this work is to bridge the gap between real-world SED and research models and facilitate a common framework that works well on both tagging and localization-level without utilizing dataset-specific knowledge.
Our contributions are:
\begin{itemize}
    \item A new, lightweight, model architecture for WSSED using L4-norm temporal subsampling.
    \item A novel thresholding technique named triple threshold, bridging the gap between tagging and localization performance.
    \item Verification of our proposed approach across three publicly available datasets, without the requirement of manually optimizing towards dataset-specific hyperparameters.
\end{itemize}

This paper is organized as follows: In \Cref{sec:approach}, we state our approach using a CRNN framework for duration robust sound event detection.
Further, in \Cref{sec:experiments}, the experimental setup is described, including our evaluation metrics.
Then in \Cref{sec:results} we provide our results and compare our approach to other models.
Additionally, we include an ablation in \Cref{sec:ablation} study, which aims to display the duration stability of our model and investigate our model's limitations.
Lastly, the paper culminates in \Cref{sec:conclusion}, where a conclusion is drawn.

\section{Approach}
\label{sec:approach}

Contrary to supervised learning approaches, where a (strong) one-to-one correspondence between an instance $x_j$ (here audio frame) and a label $\hat{y}_j$ exists, WSSED is labeled as a (weak) many-to-one relationship.
WSSED aims to distinguish between multiple events in an audio clip (multi-class), as well as separate apart co-occurring events (multi-label) at a time.
In particular, during training only an entire frame-level feature sequence $\left[ x_1, x_2, \ldots \right]$ has access to a single, clip-level label $\hat{y}(e) \in \{0,1\}, e \in [1, \ldots, E]$, where $e$ represents each individual event.
This scenario, in its essence, is multiple instance learning (MIL)~\cite{Dietterich1997}. 

It should be evident that within WSSED, localizing an event is much more complicated than tagging, since for tagging, training and evaluation criteria match, while for localization, important information such as duration is missing, meaning that a model needs to ``learn'' this information without guidance.
Due to this mismatch between training and evaluation, post-processing is vital to improve performance~\cite{Dinkel2019}.
The main idea in our work extends to the notion of duration robustness~\cite{Dinkel2019}.
We refer to duration robustness as the capability to precisely tag and, more importantly, estimate on- and off-sets of the aforementioned tagged event.
A duration robust model should also be capable of predicting on- and off-sets for both short and long duration events.

Our approach's framework can be seen in \Cref{fig:framework}, where the backbone CRNN with temporal pooling and subsampling layers learns clip-level event probabilities during training. 
For inference, we propose triple thresholding to enhance the duration robustness.

Our proposed model, which we further refer to as CDur (CRNN duration, see \Cref{tab:arch}), consists of a five-layer CNN followed by a Gated Recurrent Unit (GRU).
The model is based on the results in~\cite{Dinkel2019}, yet it is modified to enhance performance further.
Specifically, CDur utilizes double convolution blocks and three subsampling stages instead of five.
Moreover, our model experiences substantial gains by utilizing a batch-norm, convolution, activation block structure, especially when paired with Leaky rectified linear unit (LeakyReLU). 
The abstract representations extracted by the CNN front-end are then processed by a bidirectional GRU with 128 hidden units. 
%Apart from the CRNN architecture, other CDur modules will be presented in the subsequent sections.

\begin{table}[ht]
        \centering
        \caption{The CDur architecture used in this work. One block refers to an initial batch normalization, then a convolution, and lastly, a LeakyReLU (slope -0.1) activation. All convolutions use padding in order to preserve the input size. The notation $t \uparrow/\downarrow d$ represents up/down-sampling time dimension by $t$ and the frequency dimension by $d$. The model has two outputs: One clip-level, which can be updated during training, and one frame-level used for evaluation.}
        \label{tab:arch}
        \begin{tabular}{l|r|r}
            Layer & \multicolumn{2}{r}{Parameter} \\
                \hline\hline
            Block1 & \multicolumn{2}{r}{$32$ Channel, $3\times3$ Kernel} \\
            L4-Sub & \multicolumn{2}{r}{$2 \downarrow 4$} \\
                \hline
            Block2 & \multicolumn{2}{r}{$128$ Channel, $3\times3$ Kernel} \\
            Block3 &\multicolumn{2}{r}{ $128$ Channel, $3\times3$ Kernel}\\
                    L4-Sub & \multicolumn{2}{r}{$2 \downarrow 4$} \\
                \hline
                    Block4 & \multicolumn{2}{r}{$128$ Channel, $3\times3$ Kernel }\\
                    Block5 & \multicolumn{2}{r}{$128$ Channel, $3\times3$ Kernel }\\
                    L4-Sub & \multicolumn{2}{r}{$1 \downarrow 4$} \\
                    Dropout &\multicolumn{2}{r}{ $30\%$ } \\
                \hline
                    BiGRU & \multicolumn{2}{r}{$128$ Units } \\
                    Linear & \multicolumn{2}{r}{$E$ Units}  \\
                    \hline
                    & LinSoft & Upsample $4\uparrow 1$\\
            Output  & Clip-level & Frame-level \\
        \end{tabular}
\end{table}

%The entire framework can be seen in \Cref{fig:framework}.

\subsection{Temporal subsampling} 
We propose the use of subsampling towards practical, generalized sound event detection, based on the following three reasons:
1) Subsampling discourages disjoint predictions, e.g., short (frame), consecutive zero-one outputs $[0,1,0,1,\ldots] \rightarrow [0,0,0,0,\ldots]$.
2) Reduces the number of time-steps a recurrent neural network (RNN) needs to remember; therefore, it is similar to a chunking mechanism~\cite{zhai_chunking_AAAI17}, which summarizes parts of a time-sequence.
3) By subsampling in the time-frequency domain via L-norm pooling, abstract time-frequency representations can be learned by the model.

Please note that, in essence, subsampling and pooling are identical operations.
In our work, the term subsampling refers to intermediate operations within a local kernel (i.e., 2 $\times$ 2) on a spectrogram's time-frequency domain. 
In contrast, pooling refers to reducing a time variable signal to a single value (i.e., 500 probabilities $\mapsto$ 1 probability).

Subsampling is commonly done using average or max operators.
However, previous work in~\cite{Dinkel2019} showed that L-norm subsampling largely benefits duration robustness, specifically enabling the detection of short, sporadic events.
L-norm subsampling within a local kernel with size $K$ is defined as:
\[
  L_p(x) = \sqrt[p]{\sum_{x \in K} x^{p}},
\]
where, $L_p$ reduces to mean subsampling for a norm factor of $p=0$ and to max subsampling for $p=\inf$.
$L_p$-norm subsampling is an interpolation between mean and max operations, preserving temporal consistency similar to the mean operation while also extracting the most meaningful feature similar to the max operation.
In line with~\cite{Dinkel2019}, we exclusively set $p=4$ (referred to as L4-Sub) for CDur. An ablation study on the subsampling factor is also provided in \Cref{ssec:subsampling_factor}.

Subsampling is done in three stages $(s_1, s_2, s_3)$, where each stage subsamples the temporal dimension by a factor of $s_j, j \in [1,2,3]$.
The time resolution is overall subsampled by a factor of $v=4$, with the three subsampling stages being $(2, 2, 1)$.
At each of the three subsampling stages, the frequency dimension is reduced by a factor of $4$, thus after the last stage the input frequency dimension is reduced by a factor of $64$.
Additionally, a linear upsampling operation is added after the final parameter layer, which restores the original temporal input resolution $\frac{T}{v} \rightarrow T$.
We investigated utilizing upsampling during training and parameterized upsampling operations (e.g., transposed convolutions) but did not observe any performance gains.
Due to the above remarks, linear upsampling is only utilized during inference.
Therefore, CDur outputs predictions at a resolution of 50 Hz (or 20 ms/frame), enabling short sound detection.

\subsection{Temporal Pooling}
CDur outputs a frame-level probability $y_t(e) \in \left[ 0, 1 \right]$ for each input feature $x_t$ at time $t$. 
With temporal pooling, for event $e$, its event probability at clip-level $y(e) \in \left[ 0, 1 \right]$ is an aggregated probability of frame-level outputs $y(e) = \text{agg}_{\substack{t \in T}}(y_t(e))$. 
As introduced, temporal pooling is one of the significant contributors to tagging and localization performance.
In our work, we exclusively use linear softmax (LinSoft), introduced in~\cite{Wang2018} (\Cref{eq:linsoft}).

%Let $y(e) \in \left[ 0, 1 \right]$ be the aggregated clip-level event probability for event $e \in E$ from its frame-level outputs $y(e) = \text{agg}_{\substack{t \in T}}(y_t(e))$, also referred to as temporal pooling.
%From our point of view, temporal pooling is one of the significant contributors to tagging and localization performance.
%In our work, we exclusively use linear softmax (LinSoft), introduced in~\cite{Wang2018} (\Cref{eq:linsoft}).

\begin{equation}
    \label{eq:linsoft}
    y(e) = \frac{\sum_t^T y_t(e)^2}{\sum_t^T y_t(e)}
\end{equation}

In contrast to the attention pooling approach~\cite{Kong2019b,Kong2019a,lin2019specialized,Lu2018}, as well as AutoPool~\cite{McFee2018}, LinSoft is a weighted average algorithm that is not learned.
Instead, it can be interpreted as a self-weighted average algorithm.

CDur has two outputs: 1) The clip-level aggregate $y(e)$ and 2) The frame-level sequence prediction $y_t(e)$.
Since only clip-level labels are provided during training, only the clip-level aggregate $y(e)$ can be back-propagated and model parameters updated.
The frame-level output $y_t(e)$ is solely utilized for evaluation.

\subsection{Post-processing} 

Since WSSED is a multi-label multi-class classification task,
with overlapping events, post-processing is required during inference to transform a per event probability estimate $y_t(e)$ to a binary representation $\bar{y}_t(e)$, which determines whether a label is present ($\bar{y}_t(e)=1$) or absent ($\bar{y}_t(e)=0$).
It can help smoothen or enhance model performance by removing noisy outputs (e.g., single frame outputs).

We categorize post-processing algorithms into two branches: 1) Probability-level post-processing such as double and triple thresholding; 2) Hard label post-processing such as median filtering. 
Most post-processing methods, such as median filtering or double thresholding, only consider frame-level outputs for post-processing. 
In our work, we propose a novel triple thresholding technique that considers clip-level and frame-level outputs.
Though this work exclusively utilizes double and triple thresholding, we compare it with median filtering since it is the most common post-processing algorithm for WSSED.

\paragraph*{Double Threshold}

Double threshold~\cite{Dinkel2019,Kong2018} is a probability smoothing technique defined via two thresholds $\phi_{\text{hi}}, \phi_{\text{low}}$.
The algorithm first sweeps over an output probability sequence and marks all values larger than $\phi_{\text{hi}}$ as being valid predictions.
Then it enlarges the marked frames by searching for all adjacent, continuous predictions (clusters) being larger than $\phi_{\text{low}}$.
An arbitrary point in time $t$ belongs to a cluster between two time steps $t_1, t_2$ if the following holds:

\[
\text{cluster}(t) = \\
    \begin{cases}
    1, & \exists (t_1, t_2) \text{ s.t., } t_1 \leq t \leq t_2\\
    & \text{and } \forall t_0 \in [t_1, t_2], \phi_{\text{low}} \leq y_{t_0} \leq \phi_{\text{hi}}\\
    0, & \text{otherwise}.
\end{cases}
\]
The double threshold algorithm ($dt(\cdot)$) can be seen in \Cref{eq:double_threshold}.

\begin{equation}
    \label{eq:double_threshold}
    \bar{y}_t(e) = dt(y_t(e)) =
\begin{cases}
    1,& \text{if } y_t(e) > \phi_{\text{hi}}\\
    1,& \text{if } y_t(e) > \phi_{\text{low}} \\
    %  & \text{and } cluster(y_t(e), 
      &\text{and cluster}(t(e)) = 1 \\
    0,              & \text{otherwise}.
\end{cases}
\end{equation}
One potential benefit of double thresholding is that its parameters are less susceptible to duration variations and can effectively remove additional post-training optimization (e.g., window size for median filter)~\cite{Dinkel2019}.
However, double thresholding relies on robust predictions from the underlying model, since it cannot, contrary to median filtering, remove erroneous predictions (e.g., $y_t(e) >\phi_{\text{hi}}$, see \Cref{sec:ablation}).
% For all our experiments double thresholding is the default post-processing method.

%Double thresholding is a

\paragraph*{Triple Threshold}

One possible disadvantage of double thresholding is that it does not consider the clip-level output probability.
We propose triple thresholding, which extends towards double thresholding by incorporating an additional threshold on clip-level $\phi_{\text{clip}}$.
This threshold firstly removes all frame-level ($y_t(e)$) probabilities which were not predicted on clip-level ($y(e)$).
Triple thresholding is defined as \Cref{eq:triple_threshold}.
\begin{equation}
    \label{eq:triple_threshold}
%\[
    \begin{split}
    y_t(e) &= 
\begin{cases}
    y_t(e),& \text{if } y(e) > \phi_{\text{clip}} \\
    0,              & \text{otherwise}
\end{cases}
\\
\bar{y}_t(e) &= dt(y_t(e))
\end{split}
%\]
    %y(e) > \phi_{ev} 
\end{equation}

Therefore, we first remove events not predicted by the model on clip level before proceeding with double threshold post-processing.
The default $\phi_{\text{clip}}$ threshold is set to be $0.5$, since this is also the most reasonable choice when assessing tagging performance.
Experiments denoted as \textit{+Triple}, utilize $\phi_{\text{clip}}=0.5, \phi_{\text{low}} = 0.2, \phi_{\text{hi}}=0.75$.
%The threshold $\phi_{\text{clip}}=0.5$ is chosen as a reasonable def

\paragraph*{Median Filtering}

Median filtering is a conventional hard label post-processing method, meaning it is applied after a thresholding operation.
Let $Y = \left( y_1(e), \ldots, y_t(e), \ldots, y_T(e) \right)$ be a probability sequence for event $e$, $\phi_{\text{bin}}$ be a threshold and let its corresponding binary sequence be $\bar{Y} = \left( \bar{y}_1(e), \ldots, \bar{y}_t(e), \ldots, \bar{y}_T(e) \right)$ with the relation:
\[
    \bar{y}_t(e) = 
\begin{cases}
    1, & \text{if } y_t(e) > \phi_{\text{bin}} \\
  0, & \text{otherwise}.
\end{cases}
\]

A median filter then acts on the sequence $\bar{Y}$ by computing the median value within a window of size $\omega$.
In practice, a median filter will remove any event segments with frame duration $\leq \left \lfloor{\frac{\omega}{2}}\right \rfloor$ and merge two segments with distance $\leq \left \lfloor{\frac{\omega}{2}}\right \rfloor$.
In our view, median filtering skews model performance since it can delete or insert non-existing model predictions, thus making fair model comparisons unfeasible.
Many successful models utilize an event-specific median filter, where $\omega(e)$ is estimated relative to each event's duration within the labeled development set.

\section{Experimental Setup}
\label{sec:experiments}

All deep neural networks were implemented in PyTorch~\cite{PaszkePytorch}, front-end feature extraction utilized librosa~\cite{McFee2020} and data pre-processing used gnu-parallel~\cite{Tange2011a}.
Even though a plethora of front-features exist, log-Mel spectrograms are most commonly used by the SED community due to their low memory footprint (compared to spectrograms) and excellent performance~\cite{Cakir2018}.
If not further specified, all our experiments use 64-dimensional log-Mel power spectrograms (LMS) front-end features.
Each LMS sample was extracted by a 2048 point Fourier transform every 20 ms with a Hann window size of 40 ms using the librosa library~\cite{McFee2020}.
During training, zero padding to the longest sample-length within a batch is applied, whereas a batch-size of 1 is utilized during inference, meaning no padding.
A batch-size of 64 is utilized during training for all experiments.
Moreover, training uses AdamW~\cite{Loshchilov2019,KingmaB14} optimization with a starting learning rate of $\text{1e-4}$, and successive learning rate reduction if the cross-validation loss did not improve for three epochs.
Training was terminated if no loss improvement has been seen for seven epochs.
The available training data was split into a label-balanced 90\% training and a 10\% held-out validation set for model training using stratification~\cite{sechidis2011stratification}.
Furthermore, we utilized a custom sampling strategy such that each batch contained at least a single instance of each event.
Note that for all experiments, we neglected the development subset of each dataset.
All pseudo-random seeds for each experiment (model initialization, train/cv split, the order of batches) were fixed and, therefore, reproducible. The model contained overall \num[group-separator={,}]{681068} trainable parameters, having a size of 2.7 megabytes on disk, making it lightweight and possible to be deployed on embedded systems.
The source code is available.\footnote{Source code is available \url{https://github.com/RicherMans/CDur}}

\subsection{Dataset}
\label{ssec:dataset}

\begin{figure*}
    \centering
    \subfloat[URBAN-SED]{\includegraphics[height=.25\textheight,width=.33\linewidth]{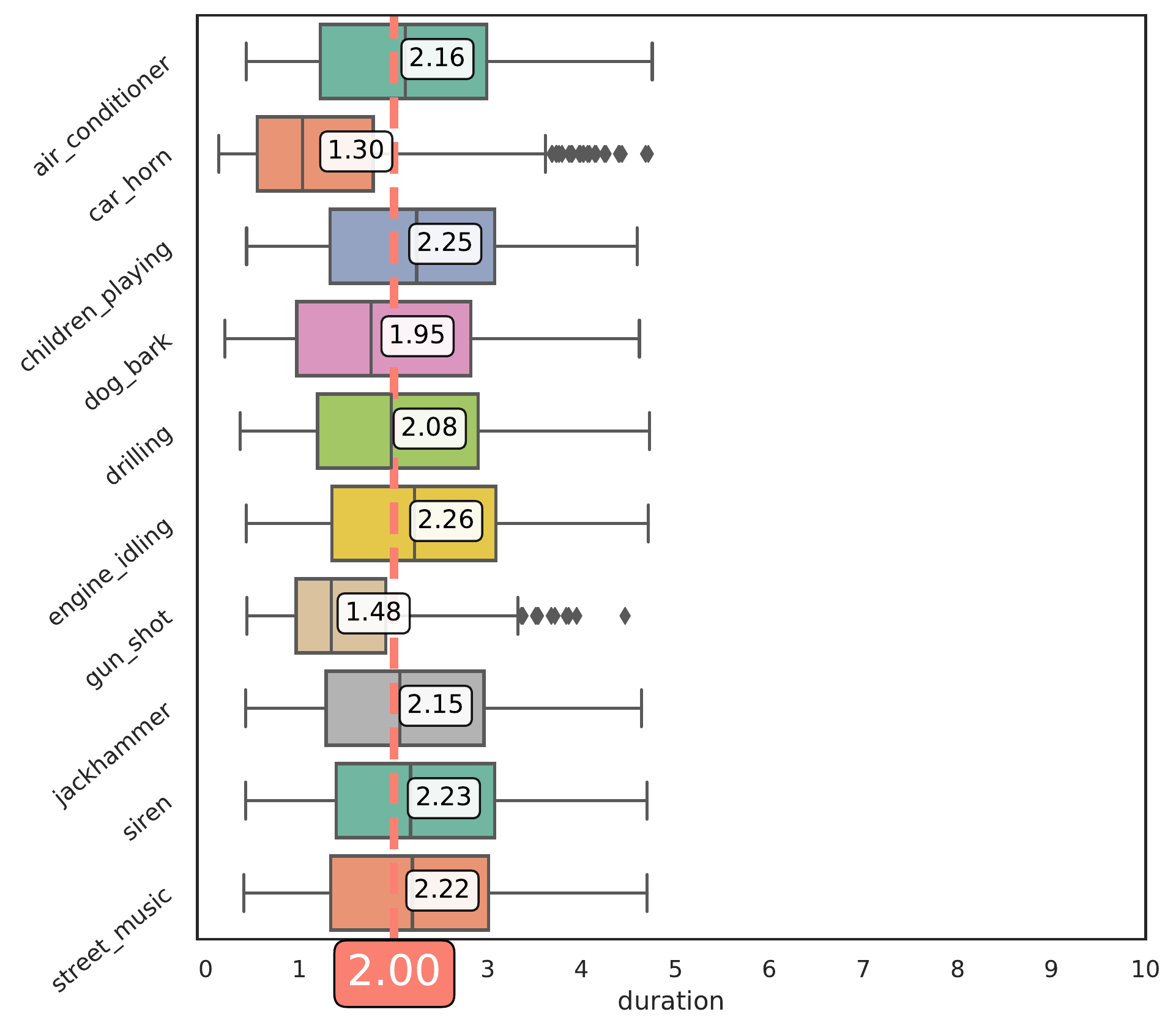}\label{fig:urban_sed_eval_data}}
    \hfill
    \subfloat[DCASE2017 Task 4]{\includegraphics[height=.25\textheight,width=.33\linewidth]{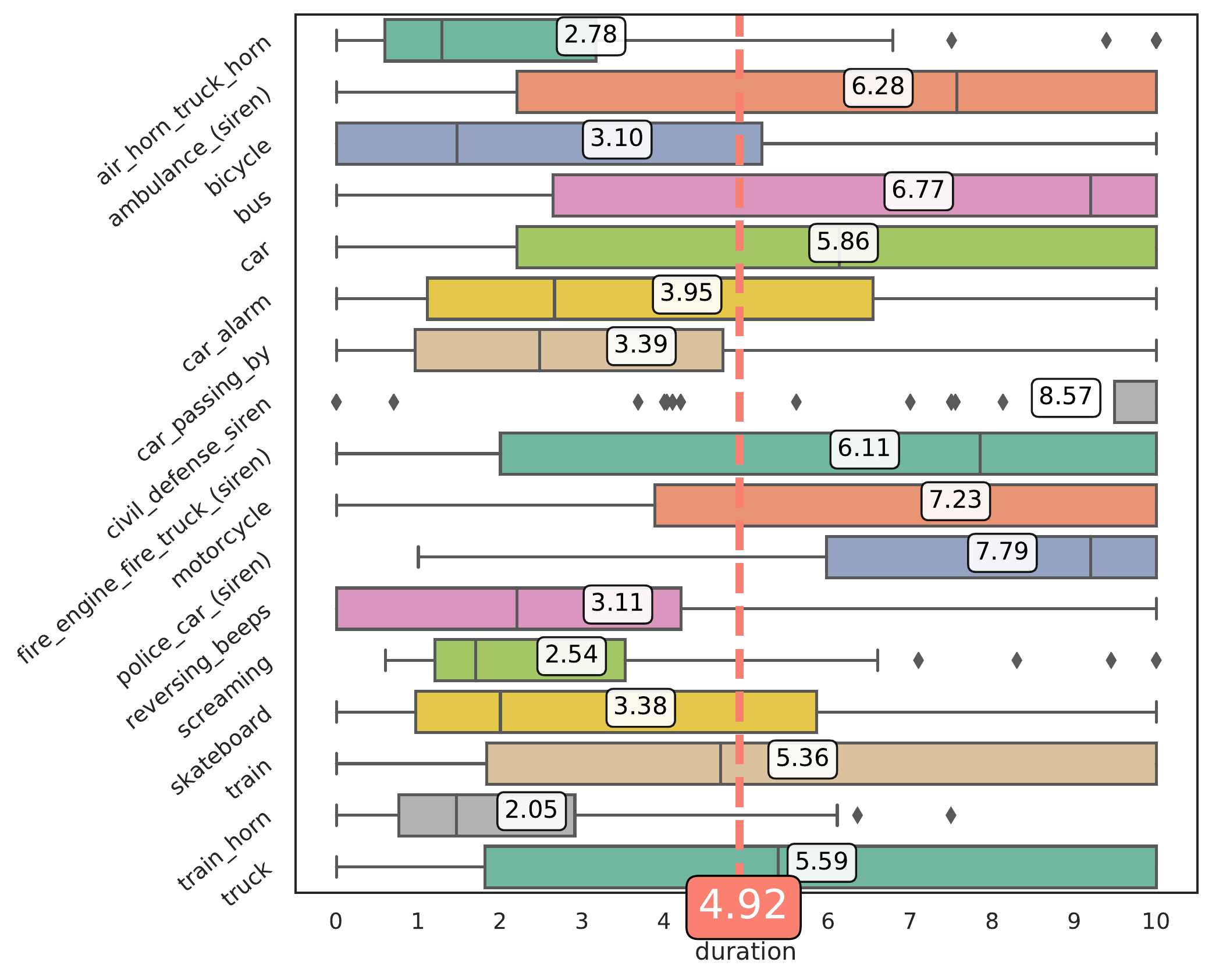}\label{fig:dcase17_eval_data}}
    \hfill
    \subfloat[DCASE2018 Task 4]{\includegraphics[height=.25\textheight,width=.33\linewidth]{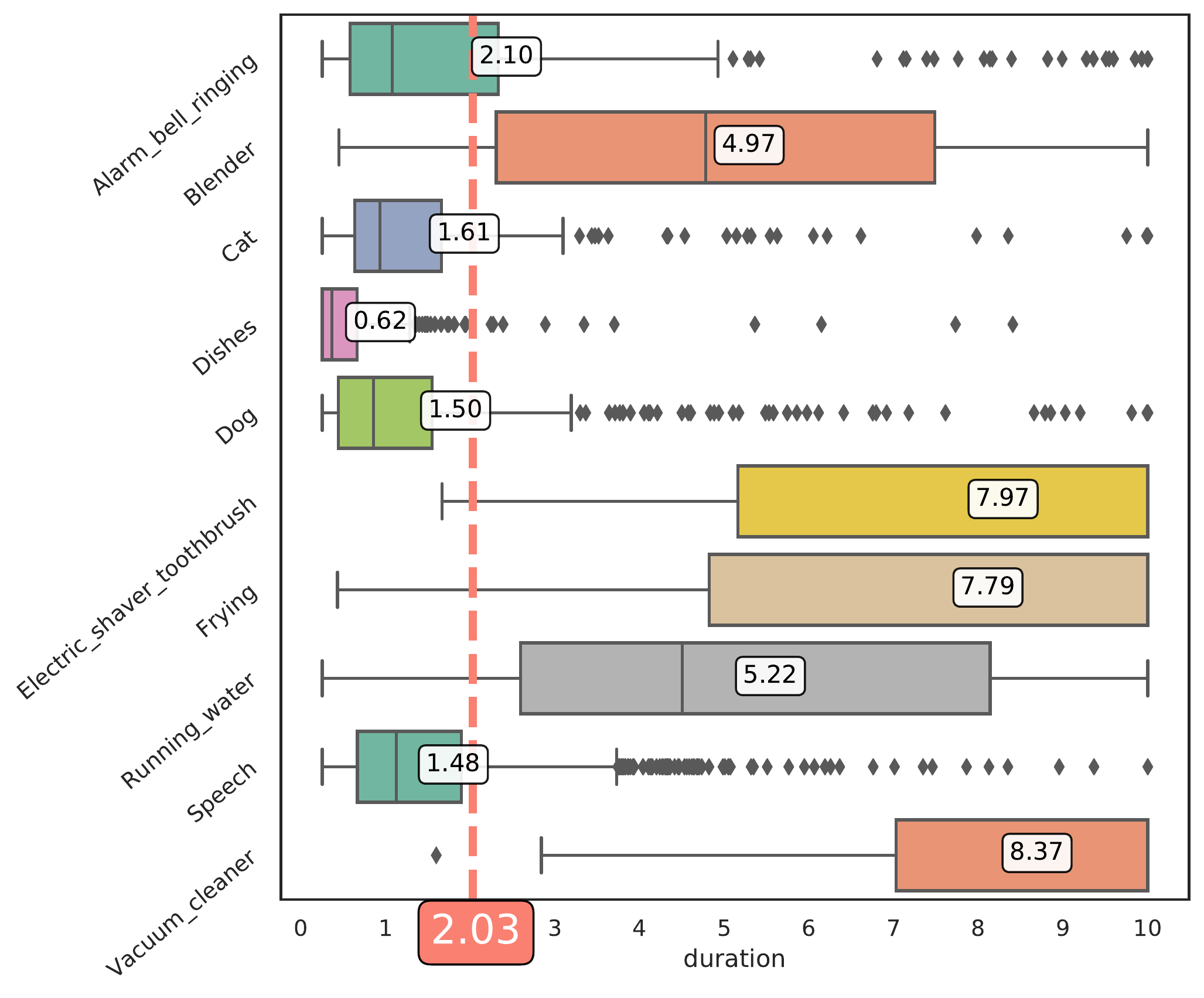}\label{fig:dcase18_eval_data}}\\
    \caption{Evaluation data duration boxplot distributions for the URBAN-SED and DCASE2017,18 datasets. Each whisker represents the minimum and maximum duration to lie within 99.3\% of a normal distribution. Each dot represents an outlier, while boxes represent the median and likely range of standard deviation within the first and third quantile. The mean duration for each event is highlighted individually (white boxes) as well as the global average (red box).}
\label{fig:all_eval_data}
\end{figure*}

This paper uses three widely researched datasets: URBAN-SED, DCASE2017 Task 4 and DCASE2018 Task 4.
Each evaluation data length distribution can be seen in \Cref{fig:all_eval_data}.
Please note that estimating events with a long duration (e.g., 10 s) is equal to audio tagging.
\begin{figure}[htpb]
    \centering
    \includegraphics[width=0.75\linewidth,clip,trim={20 10 10 10}]{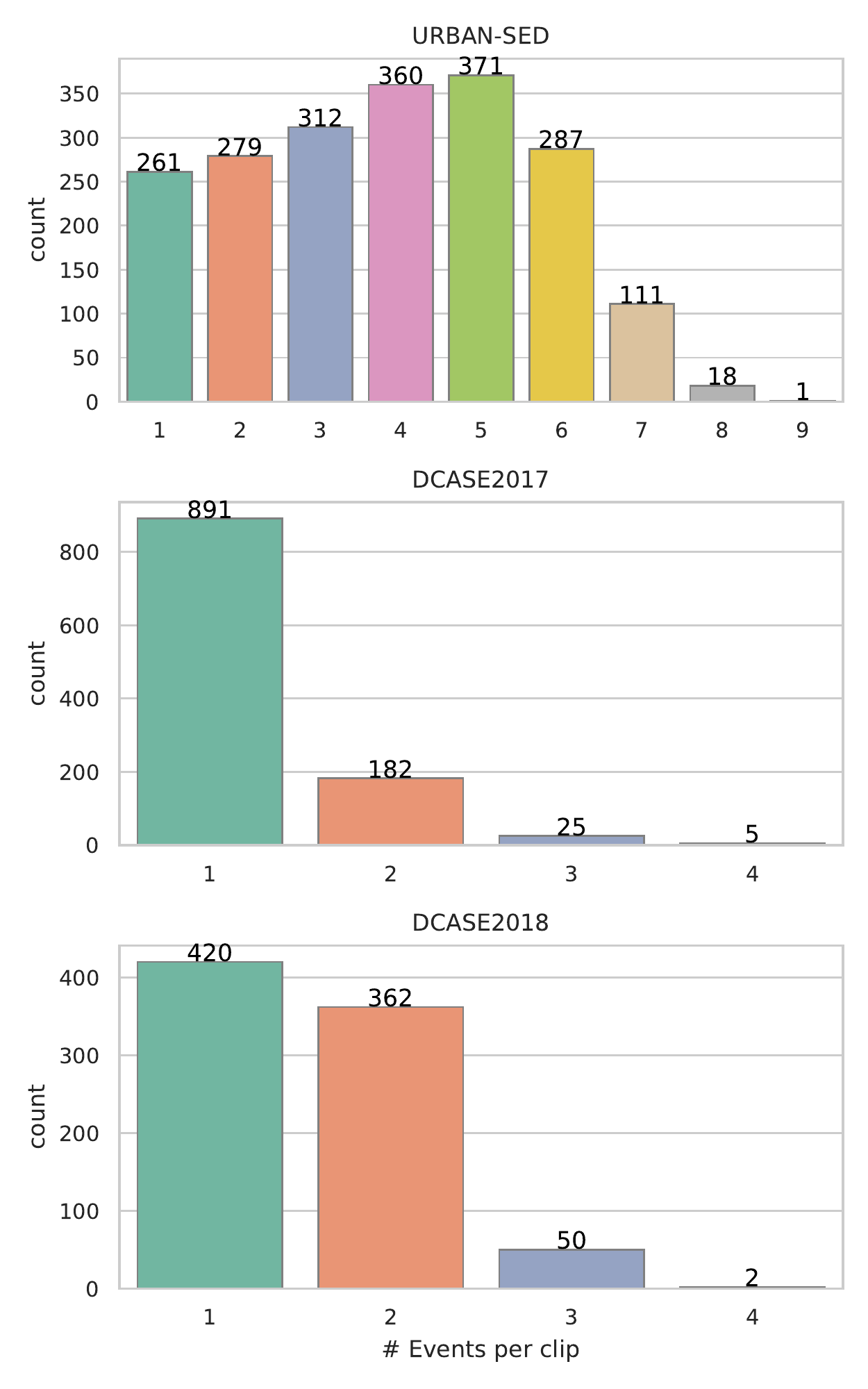}
    \caption{Number of events in a clip for each evaluation dataset.}%
    \label{fig:dataset_per_utt}
\end{figure}
We additionally provide the number of events per clip for each evaluation dataset in \Cref{fig:dataset_per_utt}.

\paragraph*{URBAN-SED} URBAN-SED~\cite{Salamon2017} is a sound event detection dataset within an urban setting, having $E=10$ event labels (see \Cref{fig:urban_sed_eval_data}).
This dataset's source material is the UrbanSound8k dataset~\cite{Salomon14_Urban8K} containing 27.8 hours of data split into 10-second clip segments.
The URBAN-SED dataset encompasses \num[group-separator={,}]{10000} soundscapes generated using the Scaper soundscape synthesis library~\cite{Salamon2017}, being split into \num[group-separator={,}]{6000} training, \num[group-separator={,}]{2000} validation and \num[group-separator={,}]{2000} evaluation clips.
The training set contains mostly 10-second excerpts, which are weakly-labeled, whereas each clip contains between one and nine events.
The evaluation dataset is strongly labeled, containing up to nine events per clip, as indicated in \Cref{fig:dataset_per_utt}.
In our work, we only utilize the training and evaluation dataset and neglect the validation one.
An essential characteristic of URBAN-SED is that since both the audio and annotations were generated computationally, the annotations are guaranteed to be correct and unbiased from human perception.
As it can be seen in \Cref{fig:urban_sed_eval_data}, the evaluation set of the URBAN-SED dataset seems to be artificially truncated due to the soundscape composition.

\paragraph*{DCASE2017} The DCASE2017 task 4 -- Large-scale weakly supervised sound event detection for smart cars dataset is also utilized.
The dataset consists of 10-second clips split into training (\num[group-separator={,}]{51172} clips), development (\num[group-separator={,}]{488} clips), and evaluation subsets (\num[group-separator={,}]{1103} clips).
Development and evaluation sets are strongly labeled, whereas the training set is weakly-labeled.
For our work, we only utilize the training and evaluation subsets and neglect the development one.
Different from URBAN-SED, DCASE2017 is subsampled from AudioSet~\cite{Gemmeke2017}, whereas $E=17$ different car-related events need to be estimated.
However, as seen from the evaluation distribution in \Cref{fig:dcase17_eval_data}, each event's duration is distributed much more naturally (between 2.5 s and 8.5 s).
In particular, the event ``civil defense siren'' seems to be the event with the most substantial duration variance, having, on average, a duration of 8.57 s, yet some samples have a duration of $<$ 1 s.
In our view, the difficulty within this dataset is the ambiguous labels, e.g., ``car'' and ``car passing by'' or the four types of sirens.
Please note that due to this dataset's naturalness, it might contain non-target events, which URBAN-SED might not.
Most clips in this dataset contain one distinct event, as indicated in \Cref{fig:dataset_per_utt}.

\paragraph*{DCASE2018} This dataset was utilized during the DCASE2018 Task4 challenge -- Large-scale weakly labeled semi-supervised sound event detection in domestic environments and contains $E=10$ unique events.
Different from the other datasets used in our work, the training data is partially unlabelled/incomplete.
While the entire training dataset consists of \num[group-separator={,}]{55990} clips, being similarly sized as the DCASE2017 dataset, only \num[group-separator={,}]{1578} clips contain labels. 
The rest \num[group-separator={,}]{54412} clips are split into in-domain data (\num[group-separator={,}]{14413} clips), where it is guaranteed that training events occur within the subset, and out-domain data (\num[group-separator={,}]{39999} clips), which might contain unknown labels.
Even though the data is drawn from AudioSet, the ten events have been manually annotated.
The evaluation data duration distribution can be seen in \Cref{fig:dcase18_eval_data}. 
Most clips in this dataset contain one or two distinct events, as indicated in \Cref{fig:dataset_per_utt}.
Compared to other datasets, DCASE2018 has the largest average difference duration-wise between the shortest (dishes, 0.62 s) and longest (vacuum cleaner, 8.37 s) events, respectively.
Moreover, it also hosts the largest variance within an event, e.g., Speech can be shorter than 1 s, but also 10 s long, indicated by a large number of outliers (\Cref{fig:dcase18_eval_data}).
Our work utilizes the training (weak) subset of the dataset for model estimation if not otherwise specified.
It should be noted that the evaluation data is identical to the DCASE2019 challenge, which, in addition to DCASE2018, adds synthetic hard labeled data for training.

\subsection{Data Augmentation}
\label{ssec:augmenatation}

A popular method in combating model overfitting and data sparsity is data augmentation.
This work investigates two different data augmentation methods, namely SpecAugment (SpecAug) and Time Shifting, in addition to a modified training criterion (Label Smoothing).
Each additional augmentation method is individually provided. 
Further, \textit{+All} refers to the utilization of all three methods described below (\textit{+SpecAug}, \textit{+LS}, \textit{+Time}).
% All results labelled as \textit{+All} utilize the following data augmentation / criteria.

\paragraph*{SpecAug}

Recently, SpecAug, a cheap yet effective data augmentation method for spectrograms, has been introduced~\cite{park2019specaugment}.
SpecAug randomly sets time-frequency regions to zero within an input log-Mel spectrogram with $D$ (here 64) number of frequency bins and $T$ frames.
Time modification is applied by masking $\gamma_{t}$ times $\eta_{t}$ consecutive time frames, where $\eta{t}$ is chosen to be uniformly distributed between $[t_0, t_0 + \eta_{t0}]$ and $t_{0}$ is uniformly distributed within $\left[0, T-\eta_{t} \right)$.
Frequency modification is applied by masking $\gamma_f$ times $\eta_{f}$ consecutive frequency bins $\left[f_0, f_0 + \eta_{f}\right)$, where $\eta_{f}$ is randomly chosen from a uniform distribution in the range of $\left[0, \eta_{f0}\right]$ and $f_0$ is uniformly chosen from the range $\left[0, D-\eta_{f}\right)$. %%$\eta_0$
For all experiments labeled \textit{+SpecAug} we set $\gamma_t = 2, \eta_{t0} = 60, \gamma_{f} = 2, \eta_{f0} = 12$.

% FREQ: 2 - 12
% TIME: 2 - 60

\paragraph*{Time Shifting}

Another beneficial augmentation method for WSSED is time-shifting.
The goal is to encourage the model to learn coherent predictions.
Effectively, given a clip of multiple audio frames $\mathbf{X}= [\mathbf{x}_1, \ldots, \mathbf{x}_T], \mathbf{X} \in \mathbb{R}^{T\times D}$, time rolling of length $\eta_{sh}$ will shift (and wrap around) the entire sequence by $\eta_{sh}$ frames to $\mathbf{X}^{\prime}= [\mathbf{x}_{\eta_{sh}}, \ldots, \mathbf{x}_{T},\ldots, \mathbf{x}_1, \ldots, \mathbf{x}_{\eta_{sh}-1}]$.
For each audio-clip, we draw $\eta_{sh}$ from a normal distribution $\mathcal{N}(0, 10)$, meaning that we randomly either shift the audio clip forward or backward by $\eta_{sh}$ frames.
Time shifting is labeled as \textit{+Time} in the experiments.

%\subsection{Training Criterion}

\paragraph*{Label Smoothing}

As our default training criterion binary cross entropy (BCE), defined as: 
\begin{equation}
    \label{eq:bce}
    \mathcal{L}(y, \hat{y}) = - \hat{y}\log(y) + (1-\hat{y})\log(1-y),
\end{equation}
between the clip-level prediction $y(e)$ and the ground truth $\hat{y}(e)$ is utilized. 
BCE encourages the model to learn the provided training ground truth labels.
However, in many real-world WSSED datasets, the ground truth labels are not necessarily correct and open to interpretation (e.g., is child babbling considered Speech?).
Since labels in WSSED are often noisy and thus unreliable, the BCE criterion (\Cref{eq:bce}) wrongly encourages overfitting towards the training ground truth labels, which might, in turn, decrease inference performance. 
Label smoothing~\cite{Szegedy2016} (LS) is a commonly used technique for regularization, which relaxes the BCE criterion to assume the ground truth itself is noisy.
Label smoothing modifies the ground truth label ($\hat{y}(e)$) for each event $e \in [ 1,\ldots, E ]$ by introducing a smoothing constant $\epsilon$, as seen in \Cref{eq:label_smoothing}.

\begin{equation}
    \label{eq:label_smoothing}
    \hat{y}_{\text{LS}}(e) = LS(\hat{y}(e)) = (1 - \epsilon) \hat{y}(e) + \epsilon \frac{1}{E}
    % LS(x) = (1 - \epsilon) x + \epsilon \frac{x}{E}
\end{equation}

In our work, we exclusively set $\epsilon=0.1$ for all experiments labeled \textit{+LS}.

\subsection{Evaluation Protocol}

Our work utilizes three evaluation metrics: 

\begin{enumerate}
    \item Audio Tagging F1 score (Tagging-F1). This metric measures the models' capability to correctly identify the presence of an event within an audio clip.
    \item Segment-F1 (Seg-F1) score. This metric is an objective measure of a given model's sound localization capability, measured by the segment-level (adjustable) overlap between ground truth and prediction. Seg-F1 cuts an audio clip into multiple fixed sized segments~\cite{Mesaros2016_MDPI}. Seg-F1 can be seen as a coarse localization metric since precise time-stamps are not required. 
    \item Event-F1 score. This metric measures on- and off-set overlap between prediction and ground truth thus is not bound to a time-resolution (like Seg-F1). The Event-F1 specifically describes a model's capability to estimate a duration (i.e., predict on- and off-set).
\end{enumerate}

Audio tagging is done by thresholding the clip-level output (after LinSoft) $y(e)$ of each event with a fixed threshold of $\phi_{\text{tag}}=0.5$ in order to obtain a many-hot vector, which is then evaluated.
The threshold $\phi_{\text{tag}}$ is fixed since this work focuses on improving Seg- and Event-F1 performance.
Segment and event F1-scores are tagging dependent, meaning that they require at least the correct prediction of an event to be assessed.
Thus, we perceive audio tagging as the least difficult metric to improve since its optimization directly correlates with the observed clip-level training criterion.
Since our work mainly focuses on duration robust estimation, Event-F1~\cite{Mesaros2016_MDPI} is used as our primary evaluation metric, which requires predictions to be smooth (contiguous) and penalizes irregular or disjoint predictions. 
To loosen the strictness of this measure, a flexible time onset (time collar, t-collar) of 200ms, as well as an offset of at most 20\% of the events` duration, is considered valid.
Furthermore, we use Seg-F1~\cite{Mesaros2016_MDPI} with a segment size of 1 second.
Lastly, in theory, two Tagging-F1 scores exist.
One can be retrieved from the frame-level prediction output (i.e., $\max_{1:T} y_t(e)$), while the other can be calculated after temporal pooling (i.e., $y(e)$).
We exclusively report the Tagging-F1 score from the clip-level output ($y(e)$) of the model in this work.
Note that the clip-level output is unaffected by post-processing.

Since each proposed F1-score metric is a summarization of individual scores, two main averaging approaches exist.
Micro scores are averaged across the number of instances (e.g., number of samples), whereas macro scores are averaged across each event (e.g., first compute F1 score on sample basis per event, then compute the average of all scores).

The sed\_eval toolbox~\cite{Mesaros2016_MDPI} is utilized for score calculation (Seg-F1 and Event-F1 scores).
Each respective dataset is evaluated using the following default evaluation metrics:
\paragraph*{DCASE2017}
The DCASE2017 challenge originally consisted of two subtasks (tagging and localization). 
The default evaluation metric on the DCASE2017 dataset is 1 second segment-level micro F1-score. 
We, therefore, report all our metrics on the micro-level (Event, Segment, Tagging).
\paragraph*{DCASE2018}
All metrics for the DCASE2018 dataset are macro-averaged, and the primary metric during the challenge was Event-F1.

\paragraph*{Urban-SED}
The default evaluation metric on this dataset is 1 second segment-level macro F1-scores, thus on average, only two segments need to be estimated for each event (see \Cref{fig:urban_sed_eval_data}).

\section{Results}
\label{sec:results}

In this section, we report and compare our results on the three publicly available datasets (see \Cref{ssec:dataset}).
If not otherwise specficied, all our reported results in this section utilize by default double thresholding with $\phi_{\text{hi}}=0.75, \phi_{\text{low}}=0.2$ and BCE (\Cref{eq:bce}) as the criterion.

\subsection{URBAN-SED}
\label{ssec:urban_sed}

In \Cref{tab:urban_sed_results}, we compare previous approaches on the URBAN-SED corpus to our CDur approach.
\begin{table}[t]
    \centering
    \caption{URBAN-SED results using our proposed CDur model. \textit{+All} refers to utilizing all augmentations. F1 scores are macro-averaged. R represents the output time resolution for a respective approach in Hz.}
    \label{tab:urban_sed_results}
    \begin{tabular}{r||llll}
        Approach  & R & Tagging-F1 & Seg-F1 & Event-F1 \\
        \hline\hline
        Base-CNN~\cite{Salamon2017} & 1  & - & 56.00  & - \\
        SoftPool~\cite{McFee2018} & 2.69 & 63.00 & 49.20 & - \\
        MaxPool~\cite{McFee2018} & 2.69 & 74.30 & 46.30 & - \\
        AutoPool~\cite{McFee2018} & 2.69 & 75.70 & 50.40 & - \\
        Multi-Branch~\cite{Huang2020} & 50 & - & 61.60 & -  \\
        Supervised SED~\cite{Martin-Morato2019} & 43.1 & - &  64.70 & -  \\
        \hline
        Ours & \multirow{6}{*}{50}  &  76.09 & 62.83 & 19.92 \\
        \textit{+LS} & & 75.00    & 61.69  & 18.74\\
        \textit{+Time} &  & 76.13 & 62.61 & 19.69\\
        \textit{+SpecAug}  & & 76.49 & 64.19 & 20.58 \\
        \textit{+All}  & & \textbf{77.13} & \textbf{64.75} & 21.73\\
        \textit{+All +Triple} & & \textbf{77.13} & \bf{64.75} & \bf{22.54}\\

    \end{tabular}
\end{table}

Our baseline CDur result can be seen to outperform all compared approaches in terms of Tagging-F1.
Moreover, its Seg-F1 score is also significantly higher (62.83\%) than all other previous WSSED approaches.
Even though CDur is capable of estimating clip- and segment-level events, it falls short of providing a competitive Event-F1 performance.
We believe that the comparatively low Event-F1 performance stems from the nature of urban auditory scenes, where most events occur in a much more random fashion compared to, e.g., domestic ones.
Moreover, the performance discrepancy between Seg-F1 and Event-F1 further exemplifies the difficulty in WSSED to obtain fine-scale onset and offset estimates.
When adding additional augmentation methods, the performance further improves (77.13 Tagging-F1, 64.75 Seg-F1).
Most notably, our augmented CDur also outperforms the fully-supervised SED system in~\cite{Martin-Morato2019} in terms of Seg-F1 (64.70\%).
Lastly, by further replacing double thresholding with triple thresholding as the post-processing method, the Event-F1 score improves from 21.73 to 22.54\%. 
\begin{figure}[htpb]
    \centering
    \includegraphics[width=\linewidth]{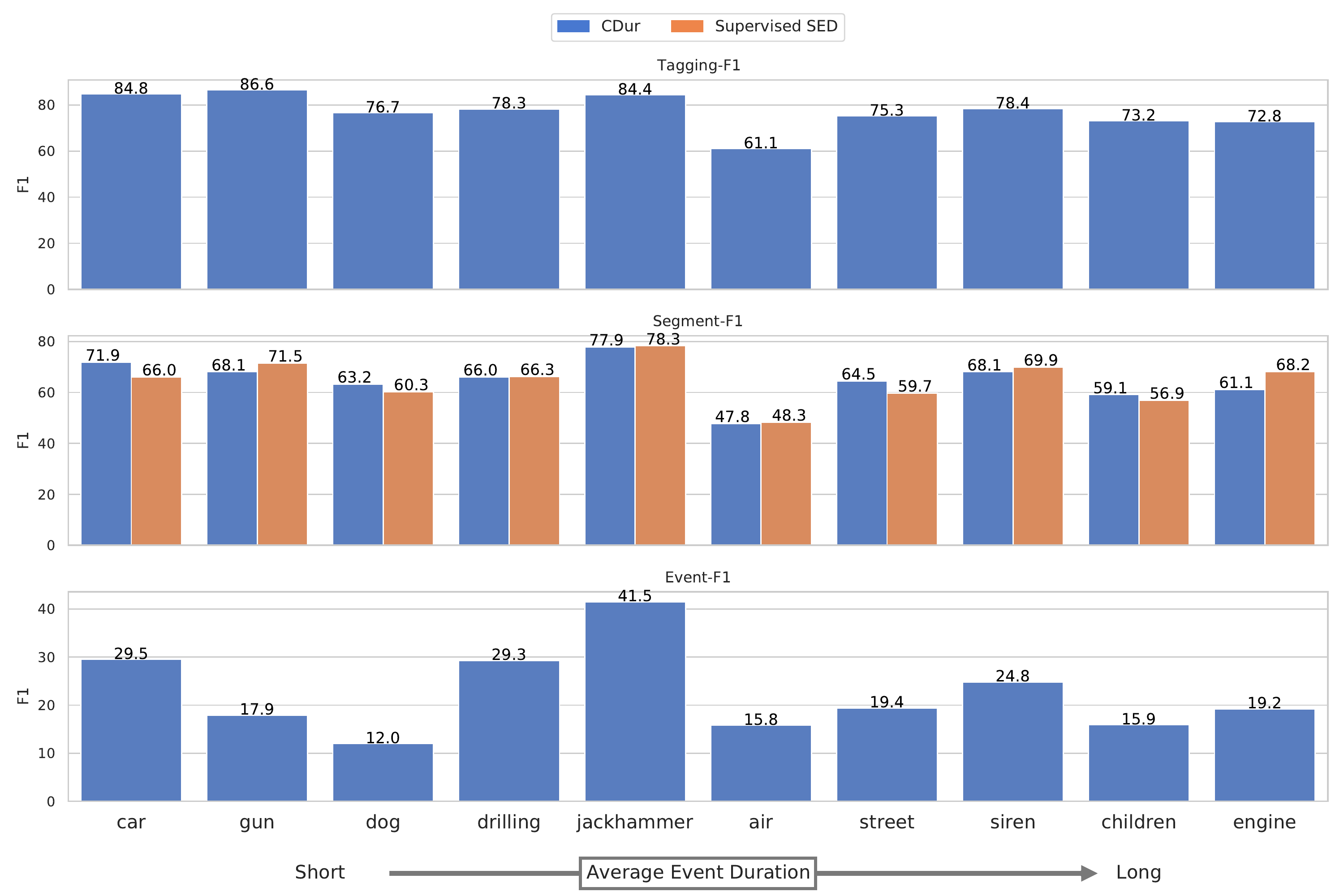}
    \caption{CDur per event F1 score results on the URBAN-SED dataset. Segment-F1 scores are compared to the next best (supervised) approach~\cite{Martin-Morato2019}.The events are sorted from (left) short to (right) long average duration.}%
    \label{fig:cdur_scores_urbansed}
\end{figure}
We provide a per event breakdown of our model performance in \Cref{fig:cdur_scores_urbansed} and compare CDur to the next best (supervised SED) model from~\cite{Martin-Morato2019}.
As seen, CDur performs evenly across all ten events in terms of Tagging-F1, averaging $\approx$ 70\% across most events.
A similar observation can also be made regarding the Seg-F1 score, where most events obtain a score of $\approx$ 65\%. 
Even though CDur only slightly outperforms the supervised approach from~\cite{Martin-Morato2019}, analyzing the per event Seg-F1 scores reveals that CDur achieves an F1 score of 71.9 compared to 66.0 on the shortest event within the dataset (car\_horn, here car).
Event-F1 results are also shown to be evenly distributed, reaching from the lowest 12.0\% for ``dog'' to 41.5\% for ``jackhammer''.
Future work is still required to exclusively estimate very short events effectively, such as in this dataset.
%We believe that the low Event-F1 scores stem from the large overlap of the dataset, hindering focused event localization.%warum ist event-F1 klein aber die anderen F1 groß?
%Due to URBAN-SED containing only short events, event-F1 scores on average low compared to the DCASE datasets.

%| Pooling   | Event F1 | Segment F1 | Tagging F1 | Paper Segment F1 | Paper Tagging F1 |
%| Soft      | 0.02     | 0.45       | 0.61       | 0.492            | 0.630            |
%| Mean      | 0.007    | 0.33       | 0.56       | 0.408            | 0.543            |
%| Max       | 0.10     | 0.51       | 0.68       | 0.463            | 0.743            |
%| Auto      | 0.06     | 0.55       | 0.69       | 0.504            | 0.757            |

\subsection{DCASE2017}
\label{ssec:dcase17}

We here provide the results on the DCASE2017 dataset, where all reported metrics are micro averaged.
\begin{table}[tb]
    \centering
\caption{DCASE2017 results compared to our approach. F1-scores are micro averaged. Best results highlighted in bold. Results underlined are fusion systems. R represents the output time resolution for a respective approach in Hz.}
    \label{tab:dcase17_results}
    \begin{tabular}{r||llll}
        Approach & R &  Tagging-F1 & Seg-F1 & Event-F1 \\
        \hline\hline
        MaxPool~\cite{McFee2018} & 2.69 & 25.70 & 25.20 & - \\
        AutoPool~\cite{McFee2018} & 2.69 & 45.40 & 42.50 & - \\ 
        Stacked CRNN~\cite{Adavanne2017} & 50 & 43.30 & 48.90 & -  \\
        Fusion GCRNN~\cite{Xu2017} & 24 & \underline{55.60} & \underline{51.80} & - \\
        GCRNN~\cite{Xu2017} & 24 & 54.20 & 47.50 & - \\
        GCCaps~\cite{Iqbal2018} & 24 & \bf{58.60} & 46.30 & - \\
        Winner SED~\cite{Lee2017b} & 1 & \underline{52.60} & \underline{\bf{55.50}} & -  \\
        \hline
        Ours & \multirow{6}{*}{50} & 52.39 & 46.12 & 15.12\\
        \textit{+LS} & & 52.07 & 46.73 & 15.60\\
        \textit{+Time} & & 49.83 & 46.60 & 16.15\\
        \textit{+SpecAug} & & 55.07 & 49.94 & 15.46\\
        \textit{+All}&  & 55.29 & 50.79 & 15.26\\
        \textit{+All +Triple} & & 55.29 & 49.93 & 15.73\\
    \end{tabular}
\end{table}
Regarding the DCASE2017 dataset results in \Cref{tab:dcase17_results}, it can be observed that even though the dataset contains the most available training data, Tagging- (52.39\%), Seg- (46.12\%), and Event-F1 (15.12\%) performance is sub-optimal.
After adding augmentation to CDur training, Tagging- (55.29\%), Seg- (50.79\%) and Event-F1 (15.26\%) performance significantly increases.
However, by replacing double threshold with triple threshold, no gains can be observed.
We believe that our performance on this dataset could improve by increasing the number of trainable parameters since the most comparative approach to our in~\cite{Xu2017} utilized at least six convolutional layers.
Moreover, note that our approach is a single model for both Tagging- and Seg-F1 evaluation, while the best comparable model~\cite{Xu2017} trained two networks with different time resolutions for each respective task.
Other approaches are specialized for one individual task (tagging, localization), e.g., the winning localization system~\cite{Lee2017b}, a large CNN fusion model, outperforms our method only regarding Seg-F1.
This is to be expected, since~\cite{Lee2017b} outputs their predictions on a coarse time resolution of at least 1 s (1 Hz), matching the segment-level criterion.
Besides, CDur is the only approach preserving acceptable performance, yet with a high time-resolution of 50 Hz. 
\begin{figure}[tpb]
    \centering
    \includegraphics[width=\linewidth]{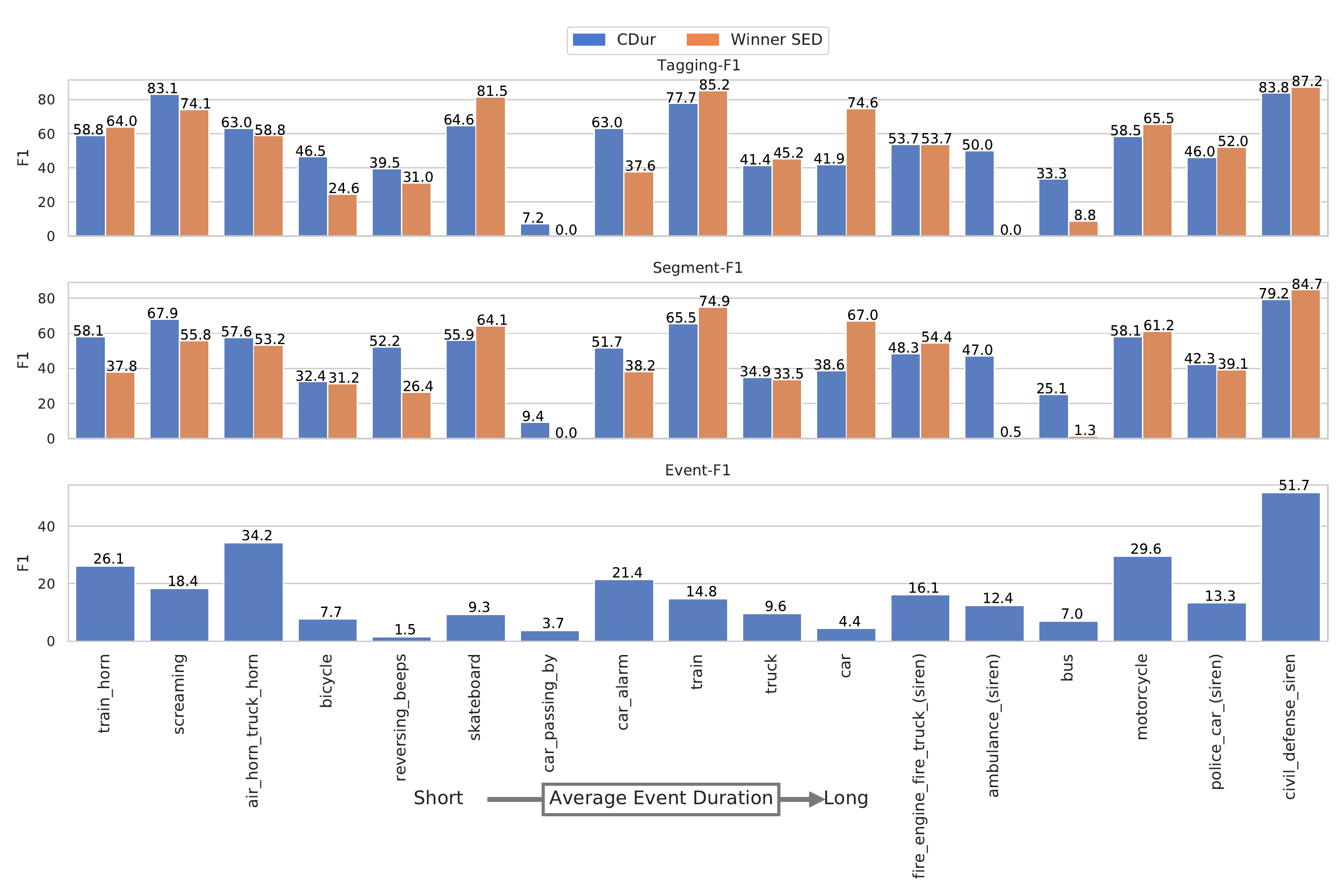}
    \caption{CDur per event evaluation F1-scores on DCASE2017. The scores are compared to the winning SED model~\cite{Lee2017b} from the DCASE2017 challenge. The events are sorted from (left) short to (right) long average duration.}%
    \label{fig:cdur_scores_dcase17}
\end{figure}
A closer look at the per event F1 scores in \Cref{fig:cdur_scores_dcase17} reveals that CDur has difficulties predicting ambiguous events such as ``car passing by'' and ``car''.
Regarding joint tagging and localization performance, CDur shows promising results for the events ``air horn'' (2.78 s), ``civil defense siren'' (8.57 s) and ``motorcycle'' (7.23 s), achieving $\approx$ 68\% Tagging-F1, $\approx$ 65\% Seg-F1 and $\approx$ 38.5\% Event-F1 average scores.
Notably, the best comparable SED model~\cite{Lee2017b} can be seen to miss some events in terms of Seg-F1.
Those events are ``car\_passing\_by'', ``ambulance'' and ``bus''. 
In stark contrast, CDur's worst Seg-F1 performance is also on ``car\_passing\_by'', but CDur still manages a Seg-F1 score of 9.4\% (against 0.0\%).
Compared with other approaches, one distinct feature of CDur is that it does not miss any event, regarding all utilized metrics.
Moreover, performance for the shortest duration events ``air horn'' (2.78 s, Event-F1 34.2\%), ``train horn'' (2.05 s, Event-F1 26.1\%), ``screaming'' (2.54 s, Event-F1 18.4\%) further exemplifies our model's capability in detecting short duration events.

\subsection{DCASE2018}

For these experiments, we utilize two given DCASE2018 subsets for model training, being the common training subset (weak) and the unlabeled in-domain subset.
Weak labels were estimated on the in-domain dataset for further training using our best performing CDur model (\textit{+All}, 70.56 Tagging-F1) by thresholding the clip-level output with a conservative value of $\phi_{\text{tag}}=0.75$.
Since not all clips obtained a label (no event has a probability higher than 0.75), we report that our predicted in-domain subset consists of 9266/14412 clips.
Note that other approaches estimating labels for the in-domain data are likely to be different from ours (such as~\cite{lin2019specialized,Liu2018,Lu2018,Kothinti2018}).
We verify that our predicted labels are usable by training CDur exclusively on the generated in-domain labels (\Cref{tab:dcase18_results}).
Further, we refer to ``Weak+'' as the merged dataset of ``Weak'' and the predicted ``In-domain'' data, containing 10844 semi-noisy clips.

\begin{table}[htpb]
    \centering
    \caption{DCASE2018 evaluation results grouped by training data. The best result is highlighted in bold, and fusion approaches are underlined.
    R represents the output time resolution for a respective approach in Hz.}
    \label{tab:dcase18_results}
    \begin{tabular}{rr||p{0.8mm}p{8mm}ll}
        Approach & Data & R & Tagging-F1 & Seg-F1 & Event-F1 \\
        \hline\hline
        Hybrid-CRNN~\cite{Kothinti2018} & Weak+ & 50 & - & - & \underline{25.40} \\
        %Multi-Scale CRNN\cite{Guo2019} & Weak & - & - & 29.20 \\ % DEVELOPMENT!
        Second'18~\cite{Liu2018} & Weak+ & 50 & - & - & \underline{29.90} \\
        Winner'18~\cite{Lu2018} & Weak+ & 16 & - & - & \underline{32.40} \\
        CRNN~\cite{Dinkel2019} & Weak & 25 & - & - & \underline{32.50}\\
        Multi-Branch~\cite{Huang2020} & Weak & 50 & - & - & 34.60 \\
        cATP-SDS~\cite{lin2019specialized} & Weak+ & 50 & 65.20 & - & 38.60 \\
        \hline\hline
        Ours & \multirow{3}{*}{Weak} & \multirow{3}{*}{50} & 69.20  & 59.89 & 31.70  \\
        \textit{+All} & & & \textbf{70.56} & 63.17 & 35.71\\
        \textit{+All +Triple} & & & 70.56 & 62.84 & 36.23\\
        \hline
        Ours & In-domain & 50 & 64.68 & 59.43 & 31.12\\
        \hline
        Ours & \multirow{6}{*}{Weak+} & \multirow{6}{*}{50} & 67.19 & 59.85 & 36.49 \\
        \textit{+LS}      &  & &     69.63 &   62.96 &     36.87 \\
        \textit{+Time}    &   & &   68.67 &   62.63 &     38.03 \\
        \textit{+SpecAug} &    & &  69.93 &   62.94 &     36.28 \\
        \textit{+All} &  & & 69.11 & \bf{63.53} & 39.18 \\
        \textit{+All +Triple} &  & & 69.11 & 63.03 & \textbf{39.42}
    \end{tabular}
\end{table}

The results in \Cref{tab:dcase18_results} indicate our models' superior performance regarding the three evaluation metrics.
CDur trained only with weak data (Event-F1 31.70) approaches performance near the winning model of 2018, which utilized additional (weak+) data.
Moreover, training our model only on its estimated, noisy labels (in-domain) leads to a similar performance when trained on the weak dataset in terms of Seg- and Event-F1.
This result shows that CDur is capable of handling noisy labels to estimate an events' duration successfully.
However, Tagging-F1 performance deteriorates from 69.20 to 64.68\% when training only on noisy labels.
Training CDur on the merged weak+ data further enhances performance in terms of Event-F1 (36.49\%) but worsens the Tagging-F1 (67.19\%) and Seg-F1 (58.12\%) performance.
% We believe that 
We believe that the additional, noisy data-enhanced onset and offset estimation accuracy, leading to an Event-F1 improvement. 
However, the inconsistency between clean (weak) and noisy (in-domain) labels possibly confused CDur's internal belief state, resulting in a Tagging-F1 performance downfall. 
Further, by explicitly modeling all labels as being noisy via label smoothing (weak+ and \textit{+All}), the Tagging-F1 performance significantly improves from 67.19 to 69.11\% and returns to original, clean label levels of only using weak data (69.20\%).
After incorporating our proposed augmentation methods as well as triple thresholding, our best Event-F1 result 39.42\% is achieved.
This result is remarkable since it would perform at eighth place on the DCASE19 challenge, which has access to hard labels during training, yet unavailable in our case.

It is notable that previously well-performing models~\cite{lin2019specialized,Huang2020,Lu2018,Liu2018} all require a fully-labeled development dataset in order to estimate a per event median filter. 
By contrast, our approach does not rely on such labels yet still achieves competitive results. 
Note that it would be possible to further enhance our performance by estimating per event thresholds, similar to~\cite{Lu2018,lin2019specialized}.
However, we refrain from doing so since one of our goals is to propose an inherently well-performing model without post-training hyperparameter tuning.
In \Cref{sec:ablation} we will further discuss this post-processing issue.
\begin{figure}[tpb]
    \centering
    \includegraphics[width=\linewidth]{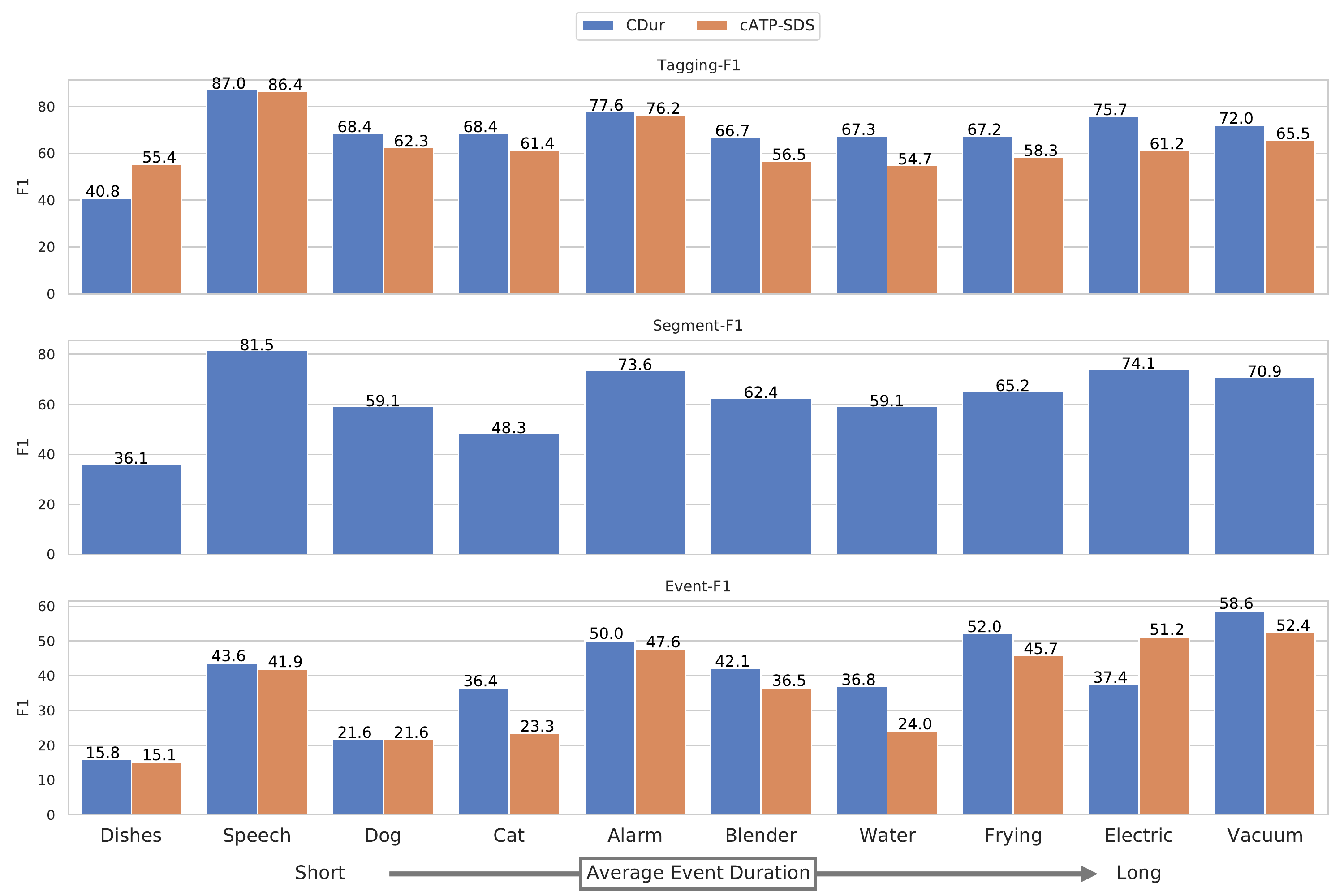}
    \caption{CDur event evaluation F1-scores on DCASE2018. The provided F1 scores are compared against the competitor (cATP-SDS) from~\cite{lin2019specialized}. The events are sorted from (left) short to (right) long average duration.}%
    \label{fig:cdur_scores_dcase18}
\end{figure}
Furthermore, the per-event F1-scores for our best performing model is shown in \Cref{fig:cdur_scores_dcase18}.
Results reveal that CDur is capable of excellent performance across all ten events, excelling at predicting long events such as ``vacuum cleaner'' (average duration 8.37 s, Event-F1 58.6\%) as well as short ones such as ``alarm'' (average duration 2.1 s, Event-F1 50.0\%).
Naturally, shorter events (cat, dog, dishes) are more challenging to predict due to their bursty nature, therefore overall, the worst-performing events for our model (1.61 s - 36.4\%,1.50 s - 21.6\%, 0.62 s - 15.8\%).
In particular, our model struggles at predicting ``dishes'' on a clip level (Tagging-F1 40.8\%), while all other labels are estimated with $\approx$ 70\% Tagging-F1 score.
When comparing CDur against cATP-SDS some interesting observations are seen.
First, only on the event ``Electric'', cATP-SDS outperforms CDur in terms of Event-F1 (37.4\% vs. 51.2\%), whereas for most other events, CDur is better.
Second, estimating short events is hard for cATP-SDS. 
This can be seen by comparing Tagging-F1 with Event-F1 scores.
Specifically, even though cATP-SDS outperforms CDur in terms of Tagging-F1 estimating ``dishes'' (40.8\% vs. 55.4\%), their Event-F1 scores on this event are near identical (15.8\% vs. 15.1\%).
This means that even though cATP-SDS is more capable of detecting the presence of this short event than CDur, it more often fails to accurately predict on- and off-sets.

\subsection{Performance influence of triple threshold}

\begin{figure*}
    \centering
    \subfloat[DCASE2017 Task 4]{\includegraphics[width=.33\linewidth]{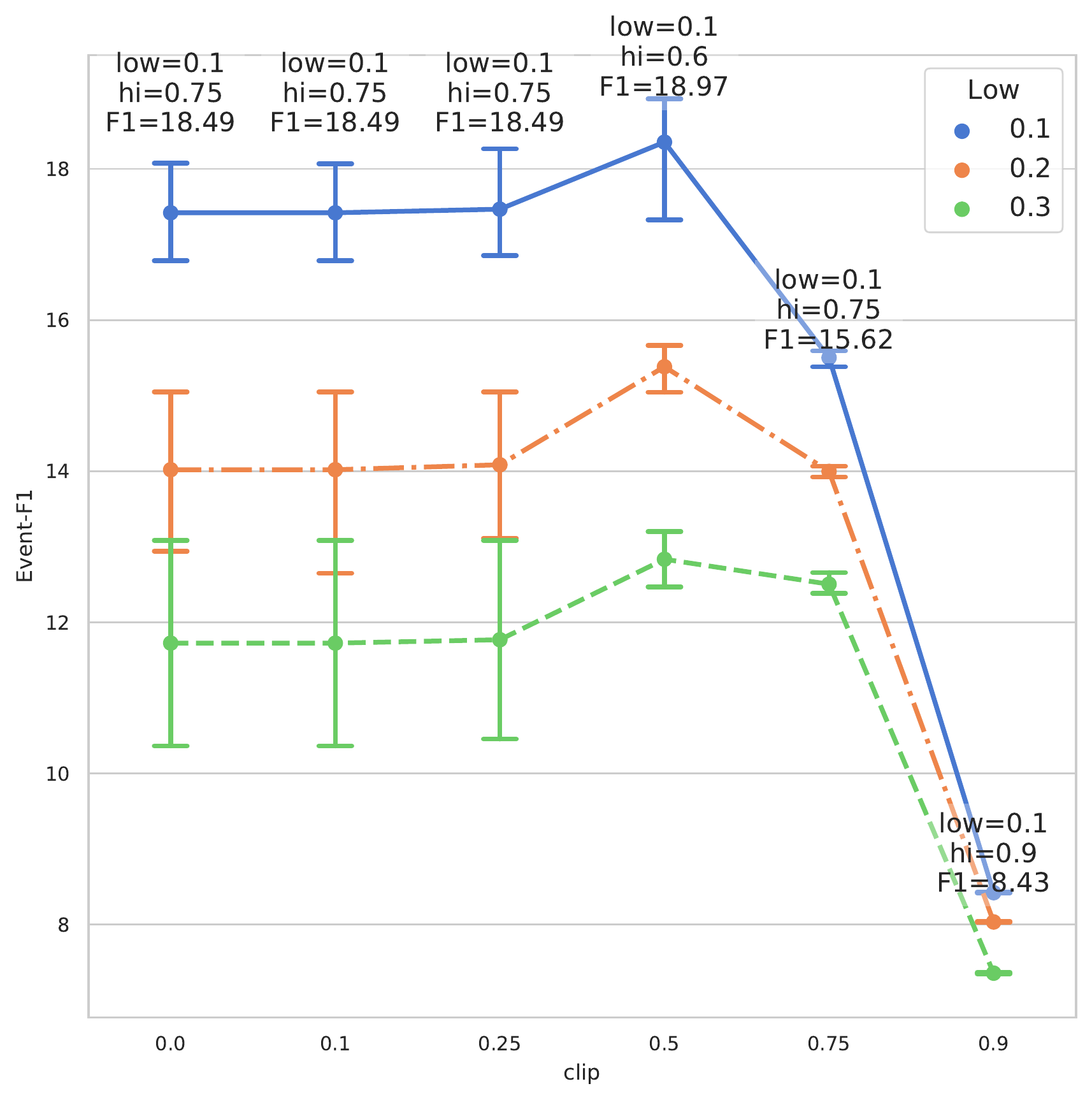}\label{fig:triple_threshold_dcase2017}}
    \hfill
    \subfloat[DCASE2018 Task 4]{\includegraphics[width=.33\linewidth]{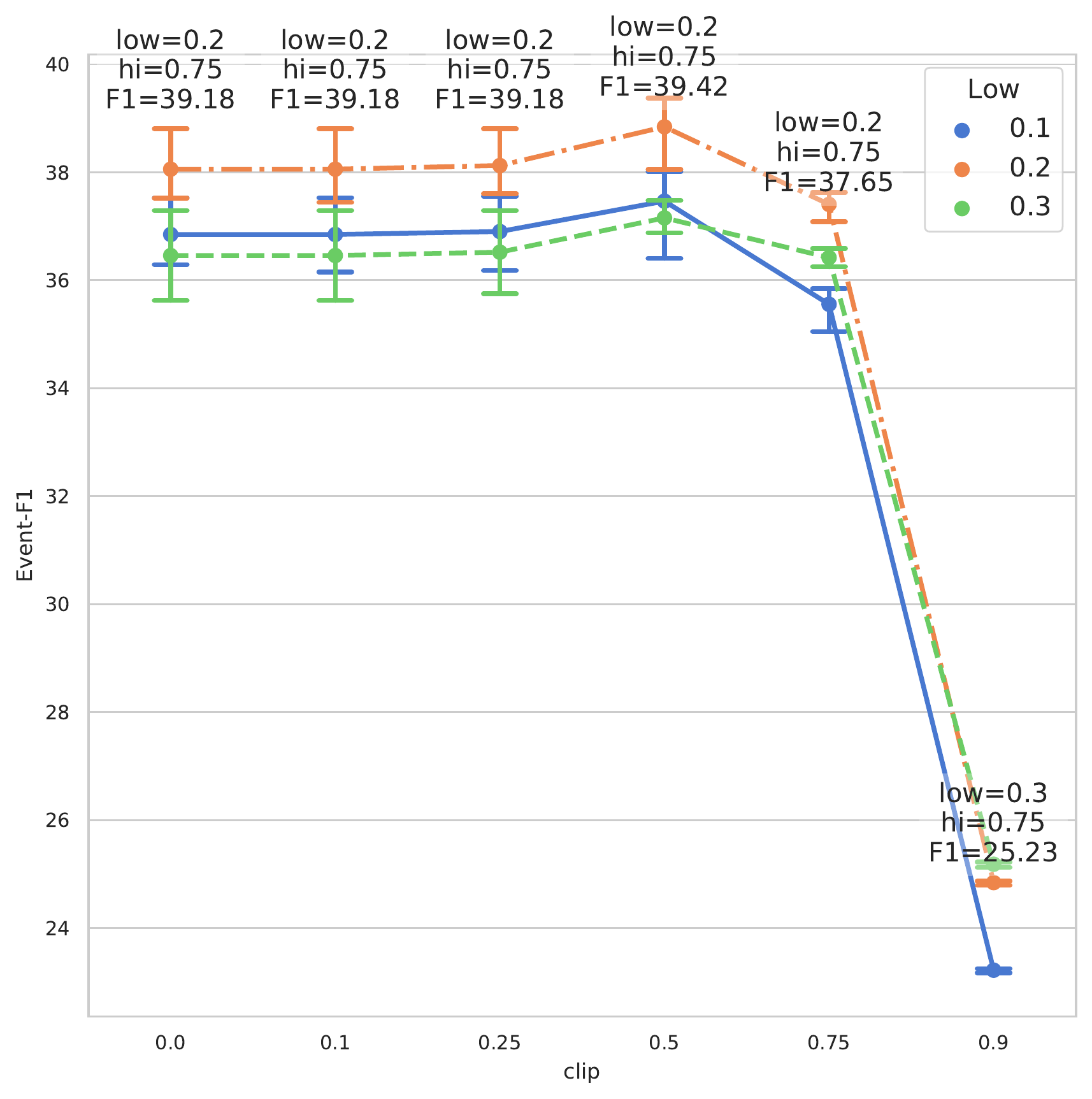}\label{fig:triple_threshold_dcase2018}}
    \hfill
    \subfloat[URBAN-SED]{\includegraphics[width=.33\linewidth]{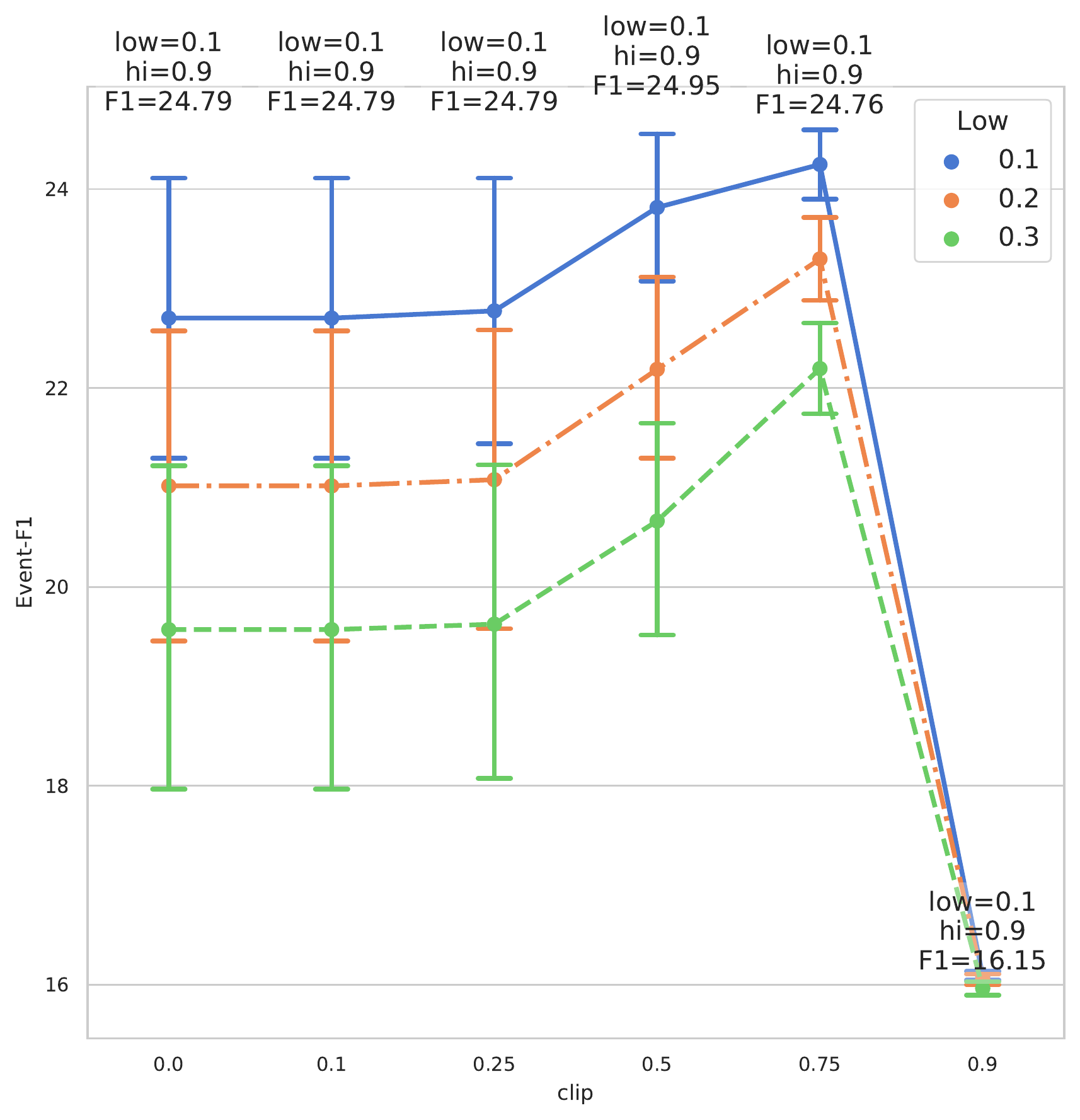}\label{fig:triple_threshold_urban_sed}}\\
    \caption{Triple threshold performance in regards to different thresholds for all three datasets. The x-axis represents the clip threshold $\phi_{\text{clip}}$, the y-axis the Event-F1 score achieved. Each color represents a lower threshold, whereas each bar represents the mean distribution of a result regarding the four utilized high thresholds $\phi_{\text{hi}} \in [0.5, 0.6, 0.75, 0.9]$. The best result for each clip-threshold is displayed. Best viewed in color.}
\label{fig:triple_threshold_investigation_all}
\end{figure*}

Another essential question to ask is how the triple threshold behaves when utilizing different thresholds and if it indeed improves performance.
Recall that triple thresholding with $\phi_{\text{clip}}=0$ reduces to double thresholding (\Cref{eq:double_threshold}).
Here we investigate the following thresholds: $\phi_{\text{low}} \in [0.1, 0.2, 0.3], \phi_{\text{hi}} \in [0.5, 0.6, 0.75, 0.9], \phi_{\text{clip}} \in [0.0, 0.1, 0.25, 0.5, 0.75, 0.9]$.
The results for each respective dataset in terms of Event-F1 can be seen in \Cref{fig:triple_threshold_investigation_all}.

Our proposed setting with $\phi_{\text{clip}}=0.5$ and $\phi_{\text{hi}}=0.75, \phi_{\text{low}}=0.2$ can be seen to indeed perform best on the DCASE2018 dataset (see \Cref{fig:triple_threshold_dcase2018}).
However, for the other two datasets, a lower threshold of $\phi_{\text{low}}=0.1$ seem to be favorable on the DCASE2017 and URBAN-SED datasets, culminating in Event-F1 scores of 18.97 (DCASE2017, micro) and 24.95 (URBAN-SED, macro) respectively.
Again, please note that considerable gains can be obtained by optimizing our thresholds if we choose to optimize towards a specific dataset.
For example, our best reported DCASE2017 model (micro Event-F1 15.73) can be improved to 18.97, by modifying $\phi_{\text{low}}=0.1$.
The same observation can be made for URBAN-SED (22.54 $\rightarrow$ 24.95).

A clip threshold of $\phi_{\text{clip}}=0.5$ seems to be a valid choice within our investigated thresholds since that threshold performs best on the DCASE2017/2018 datasets.
On the URBAN-SED dataset, larger thresholds with a high variance should be preferred ($\phi_{\text{low}}=0.1,\phi_{\text{hi}}=0.9,\phi_{\text{clip}}=0.75$).
However, still the best performing approach (24.95 Event-F1) on the URBAN-SED dataset utilizes $\phi_{\text{clip}} = 0.5$.
Correctly, by observing the trend of $\phi_{\text{clip}}$ from left to right (left = double threshold $\phi_{\text{clip}}=0.0$), one can observe that triple thresholding is effective on all three datasets and improves the average performance up until $\phi_{\text{clip}} = 0.9$ (see straight lines in \Cref{fig:triple_threshold_investigation_all}).

% Cat, Dishes are 
% First, the short event Cat 

\section{Ablation}
\label{sec:ablation}
    
%In this ablation study we focus on the impact of post-processing towards final event and segment level performance.

\subsection{Temporal pooling alternatives to LinSoft}
\label{ssec:tempoal_pooling_ablation}

This ablation study investigates the influence of CDur's performance in regards to its temporal pooling layer.
The default temporal pooling layer (LinSoft) is replaced by four commonly utilized pooling functions, described in \Cref{tab:temporal_pooling_formulation}.
Soft- and Auto pooling have been introduced in~\cite{McFee2018}.
Note that Auto pooling~\cite{McFee2018} learns a non-constrained weight parameter $\alpha(e) \in \mathbb{R}$ for each respective event $e$ (initialized as one), whereas Attention pooling~\cite{Wang2018} uses a per timestep, per event weight $w_t(e) \in [0, 1]$.

\begin{table}[htbp]
    \centering
    \caption{Temporal pooling alternatives to LinSoft investigated in this work. Soft- and Max pooling functions are parameter-free, while Auto and Attention pooling are parameterized.}
    \label{tab:temporal_pooling_formulation}
    \begin{tabular}{r||l}
        Temporal pooling & Formulation \\
        \hline\hline
        Soft & $y(e) = \sum_{t}^T y_t(e) \frac{\exp{y_t(e)}}{\sum_j^T \exp{y_j(e)}}$\\
        Max & $y(e) = \max\limits_{t} y_t(e) $\\
        Auto & $y(e) = \sum_{t}^T y_t(e) \frac{\exp{\left(\alpha(e) y_t(e)\right)}}{\sum_j^T \exp{\left(\alpha(e) y_j(e)\right)}}$\\
        Attention & $y(e) = \frac{\sum_{t}^T w_t(e) y_t(e)}{\sum_{j}^T w_j(e)}  $
    \end{tabular}
\end{table}

The ablation study results can be seen in \Cref{tab:ablation_tempoal_pooling}.
The results show that, indeed, LinSoft is to be seen as the best performing pooling method for CDur. 
In the case of DCASE2017, Auto and Soft pooling are both the closest competitors to LinSoft.
However, on the URBAN-SED dataset, Soft and Auto pooling both completely fail to generate accurate on and offsets, indicated by an Event-F1 of 0.0.
Further, while Attention shows competitive performance across all datasets in terms of all metrics, it is still mostly inferior to LinSoft, specifically in terms of Event-F1.
Most importantly, across all three proposed datasets, LinSoft is the only temporal pooling method that provides excellent performance, e.g., Auto and Soft pooling outperform Attention on the DCASE2017 dataset but are mainly inferior to Attention on the DCASE2018 and URBAN-SED datasets. 
% Lastly, the naive Max pooling operation can still yield 

\begin{table}[htbp]
\caption{Comparison of different temporal pooling functions compared to LinSoft. DCASE2018 results are trained on the weak dataset. Post-processing is set to the default double threshold, and no augmentation is applied.}
\label{tab:ablation_tempoal_pooling}
\centering
\begin{tabular}{rr||lll}
Task &      Pooling &    Tagging-F1 &    Seg-F1 &    Event-F1 \\
\hline\hline
\multirow{5}{*}{DCASE2017}   &      LinSoft &         \textbf{52.39} &         \textbf{46.12} &       \textbf{15.12} \\
   &         Auto &         47.53 &         44.25 &       13.06 \\
   &         Soft &         50.45 &         44.18 &       13.19 \\
   &    Attention &         51.04 &         41.63 &        8.99 \\
   &          Max &         47.63 &         32.34 &        7.39 \\
   \hline
 \multirow{5}{*}{DCASE2018}   &      LinSoft &         \textbf{69.20} &         \textbf{59.89} &       \textbf{31.70} \\
   &         Auto &         67.13 &         58.58 &       19.90 \\
   &         Soft &         68.01 &         56.76 &       17.79 \\
   &    Attention &         66.52 &         55.85 &       21.95 \\
   &          Max &         66.01 &         45.80 &       17.85 \\
   \hline
\multirow{5}{*}{URBAN-SED}    &      LinSoft &         \textbf{76.09} &         \textbf{62.83} &       \textbf{19.92} \\
    &         Auto &         75.14 &         38.67 &        0.00 \\
    &         Soft &         73.87 &         37.47 &        0.00 \\
    &    Attention &         74.61 &         59.89 &       19.33 \\
    &          Max &         74.50 &         56.70 &       16.19 \\
% ------------ &  ----------- &  ------------ &  ------------ &  ---------- \\
% DCASE 2018   &      LinSoft &         67.19 &         59.85 &       36.49 \\
% DCASE 2018   &         Auto &         67.10 &         56.76 &       16.15 \\
% DCASE 2018   &         Soft &         67.66 &         56.73 &       16.56 \\
% DCASE 2018   &    Attention &         68.79 &         59.66 &       18.71 \\
% DCASE 2018   &          Max &         67.03 &         57.29 &       20.08 \\

\end{tabular}
\end{table}

\subsection{Subsampling factor influence}
\label{ssec:subsampling_factor}

Another ablation study in our work focuses on the subsampling factor $v$ (default $v=4$) and its implications towards duration robustness and Event-F1 performance.
Here we compare $v=4 \mapsto (2,2,1)$, against three other factors ($v=1 \mapsto (1,1,1), v=2 \mapsto (2,1,1), v=8 \mapsto (2,2,2)$).
The temporal subsampling function is the default L4-sub (see \Cref{tab:arch}).
The result of this experiment can be seen in \Cref{tab:subsampling_ablation}.
\begin{table}[tb]
\centering
\caption{Influence of different subsampling factors $v$ on all three datasets. Our default subsampling factor in our work is 4. The DCASE2018 model is trained on the weak training set. Best result for each respective metric is highlighted in bold.}
\label{tab:subsampling_ablation}
\begin{tabular}{rr||lll}
Task &  Factor $v$ &  Tagging-F1 &  Seg-F1 &  Event-F1 \\
\hline\hline
\multirow{4}{*}{DCASE2017} &       1 &       50.33 &       45.95 &      9.05 \\
&       2 &       44.50 &       40.74 &     11.82 \\
&       4 &       52.39 &       46.12 &     \textbf{15.12} \\
&       8 &       \textbf{52.72} &       \textbf{47.22} &     14.79 \\
\hline
\multirow{4}{*}{DCASE2018} &       1 &       \textbf{69.95} &       \textbf{60.80} &     26.71 \\
 &       2 &       69.75 &       60.16 &     26.88 \\
 &       4 &       69.20 &       59.89 &     31.70 \\
 &       8 &       66.47 &       58.15 &     \textbf{33.18} \\
% \multirow{4}{*}{DCASE2018} &       1 &       68.52 &       61.78 &     29.79 \\
%  &       2 &       66.68 &       60.07 &     28.44 \\
%  &       4 &       67.19 &       59.85 &     \textbf{36.49} \\
%  &       8 &       \textbf{67.56} &       \textbf{61.85} &     35.36 \\
 \hline
\multirow{4}{*}{URBAN-SED}  &       1 &       74.29 &       60.26 &     11.86 \\
&       2 &       74.41 &       60.13 &     14.70 \\
&       4 &       \textbf{76.09} &       \textbf{62.83} &     19.92 \\
&       8 &       75.00 &       61.18 &     \textbf{20.83} \\
\end{tabular}
\end{table}
For all experiments, a clear trend can be observed that a subsampling factor of one or two perform worse than factors 4 and 8.
A larger subsampling factor of 8 is seen to benefit Tagging- and Seg-F1 performance on DCASE2017/8 datasets.
These results are in line with our previous observations in~\cite{Dinkel2019}. 
The default subsampling factor of 4, while not always performing better than a factor of 8, can be seen to be a trade-off between tagging and localization performance.
Also, even though a subsampling factor of 8 improves Event-F1 performance on the DCASE2018 and URBAN-SED datasets, we believe that a subsampling factor of 4 should be preferred on any unknown dataset due to its robustness to possibly unknown, short events.
When comparing the shortest events for each dataset according to the subsampling factor in \Cref{tab:short_events_results}, it can be seen that a factor of 4 is overall the most robust choice on all three datasets.
\begin{table}[tb]
\centering
\caption{Event-F1 scores for the shortest event in each dataset, being Dishes (DCASE2018), Car\_horn (URBAN-SED) and Train\_horn (DCASE2017).}
\label{tab:short_events_results}
\begin{tabular}{r||lll}
Subsampling factor $v$ &  Dishes &  Car\_horn &  Train\_horn \\
\hline
Avg. Duration & 0.62 s & 1.30 s & 2.05 s \\
\hline\hline
1       &    13.3 &      21.8 &        23.1 \\
2       &    16.0 &      31.8 &        24.9 \\
4       &    \textbf{23.2} &      32.0 &        \textbf{28.7} \\
8       &    16.2 &      \textbf{33.9} &        21.7 \\
\end{tabular}
\end{table}
Optimizing the subsampling factor can yield significant performance gains in terms of Event-F1. However, this optimization requires prior knowledge about each event's duration, which might not be available.

\subsection{Model comparison without post-processing}
\label{ssec:post_processing_ablation}

Another critical question is whether our model is inherently capable of producing duration robust (i.e., high Event-F1) predictions or if this is solely due to the applied post-processing approach. 
Thus, this ablation study focuses on removing any post-processing from a trained model (further denoted as \textit{-Post}) and using binary thresholding $y_t(e)>\phi_{\text{bin}}, \phi_{\text{bin}} = 0.5$ to re-evaluate frame-level predictions.
Since most works optimize their post-processing methods towards the given dataset, performance and generalization capability can be skewed towards that specific dataset.
We provide this ablation study for two successful models on the URBAN-SED and DCASE2018 datasets, respectively.
Note that we do not compare on the DCASE2017 dataset since our proposed model does not outperform contemporary approaches. 
Moreover, the main objective in the DCASE2017 challenge was to predict on a 1-second scale, meaning that the ablation would compare rough-scale estimates (1 s) with our fine-scale CDur (20 ms), which we believe does not provide further insights into CDur.
Thus, we reimplemented the currently best performing model~\cite{lin2019specialized} on the DCASE2018 dataset as well as as~\cite{McFee2018} on the URBAN-SED for comparison of duration robustness.

\paragraph*{DCASE2018}

As can be seen in \Cref{tab:ablation}, our reimplemented cATP-SDS model (34.07 Event-F1) is within the scope of their reported average performance~\cite{lin2019specialized} as well as better performing in terms of Tagging-F1.
\begin{table}[htpb]
    \centering
    \caption{Post-processing ablation (\emph{-Post}) results comparing cATP-SDS with CDur on the DCASE2018 evaluation dataset.}
    \label{tab:ablation}
    \begin{tabular}{rr||lll}
        Approach & Data & Tagging-F1 & Seg-F1 & Event-F1 \\
        \hline\hline
        cATP-SDS~\cite{lin2019specialized} & Weak+ &  65.20 & - & 38.60 \\
        \hline
        Our cATP-SDS~\cite{lin2019specialized} & Weak & 58.98 & 55.20 & 28.65\\
        \emph{-Post} & Weak & 58.98 &  17.83 & 3.77\\
        Our cATP-SDS~\cite{lin2019specialized} & Weak+ & 66.11 & 60.39 & 34.07 \\
        \emph{-Post} &  Weak+ & 66.11 & 17.74 & 2.74 \\
        %Re + Aug & Weak+& 65.91 & 61.79 & 39.54\\
        \hline
        Ours & Weak & 69.20  & 59.89 & 31.70  \\
        \emph{-Post} & Weak & 69.20 & 62.50 & 23.57 \\
        \hline
        Ours (Best) & Weak+& 69.11 & 63.03 & 39.42 \\
        \emph{-Post} & Weak+& 69.11 & 63.97 & 27.05 \\
    
    \end{tabular}
\end{table}
Our early stopping and balanced batch sampling training schedule, as well as possibly better noisy in-domain labels, could be the explanation behind this performance improvement.
It is also indicated that the choice of post-processing approach is crucial to event-level performance. 
However, post-processing has little effect on segment-level performance.
CDur's performance is affected by the choice of post-processing method (here double/triple thresholding).
In absolute, our model drops as much as 12\% in Event-F1 score when removing double/triple thresholding.
However, even though event-level performance is negatively affected, segment-level performance is enhanced for both our approaches.
This is likely due to default post-processing method using a conservative threshold choice of $\phi_{\text{hi}} = 0.75$ (Seg-Precision 74.51\%, Seg-Recall 56.80\%), while $\phi_{\text{bin}} = 0.5$ enhances recall performance (Seg-Precision 73.70\%, Seg-Recall 58.35\%), thus improving Seg-F1.
Conservative threshold values have also been reported to impact Event-F1 performance~\cite{Bilen2019} positively.
Therefore, we can observe an absolute increase from 0.9 to 3.0\% Seg-F1 for our models.
In contrast to our approach, cATP-SDS~\cite{lin2019specialized} seems to be strongly dependent on the post-processing method.
Its event and segment level performance plummets for both weak (28.65 $\rightarrow$ 3.77) and weak+ (34.07 $\rightarrow$ 2.74) training sets.
The reason for this behavior is their dependence on clip-level post-processing, which effectively filters out false-positives.
Different from cATP-SDS, CDur does not require post-processing in order to effectively localize sounds and estimate each sound's respective duration.

%Further, we prove sample outputs of both models to 

%since activation thresholds between our chosen postprocessing methods ($0.75 \rightarrow 0.5$) conser

\paragraph*{URBAN-SED}

Since most previous work, such as~\cite{McFee2018} only utilized segment-level micro F1 scores. We reimplemented those models to compare the Event-F1 performance to ours. 
We also utilized the same training schema (Pitch augmentation, binarization post-processing) as in~\cite{McFee2018}.
These results show that our reimplementations perform worse in terms of Tagging-F1 performance, but the Seg-F1 score improves compared to the original publication (see \Cref{tab:urban_sed_results}).
It can be expected that an improved Seg-F1 can correlate with an improved Event-F1.

\begin{table}[htpb]
    \centering
    \caption{Post-processing ablation (\emph{-Post}) results comparing~\cite{McFee2018} (no post-processing) approaches with CDur on URBAN-SED. Models with a ``Our'' prefix are reimplementations for metric comparison purposes.}
    \label{tab:ablation_urban_sed_results}
    \begin{tabular}{r||lll}
        Approach  & Tagging-F1 & Seg-F1 & Event-F1 \\
        \hline\hline
        SoftPool CNN~\cite{McFee2018}  & 63.00 & 49.20 & - \\
        MaxPool CNN~\cite{McFee2018}  & 74.30 & 46.30 & - \\
        AutoPool CNN~\cite{McFee2018} & 75.70 & 50.40 & - \\
        \hline
        Our SoftPool CNN~\cite{McFee2018}  & 59.86 & 45.80 & 2.45 \\
        Our MaxPool CNN~\cite{McFee2018}  & 74.34 & 58.76 & 13.04 \\
        Our AutoPool CNN~\cite{McFee2018} & 72.74 & 56.68 & 6.89 \\
        \hline
        Ours (Best)  &  77.13 & 64.75 & 21.73  \\
        \emph{-Post} & 77.13 & 63.86 & 14.54\\
    \end{tabular}
\end{table}

Please note that all models reported in \cite{McFee2018} only utilize a binary threshold strategy, identical to our (-Post) result, as post-processing.
By removing our post-processing, we can compare both approaches from a common point of view.
As it can be seen in \Cref{tab:ablation_urban_sed_results}, CDur is capable of outperforming the best localization model (MaxPool CNN), in terms of segment and Event-F1, as well as capable of outperforming the best tagging model (AutoPool CNN).

\subsection{Quantitative results}

\begin{figure}[htp]
    \centering
    \includegraphics[width=0.95\linewidth,trim={20 50 50 70},clip]{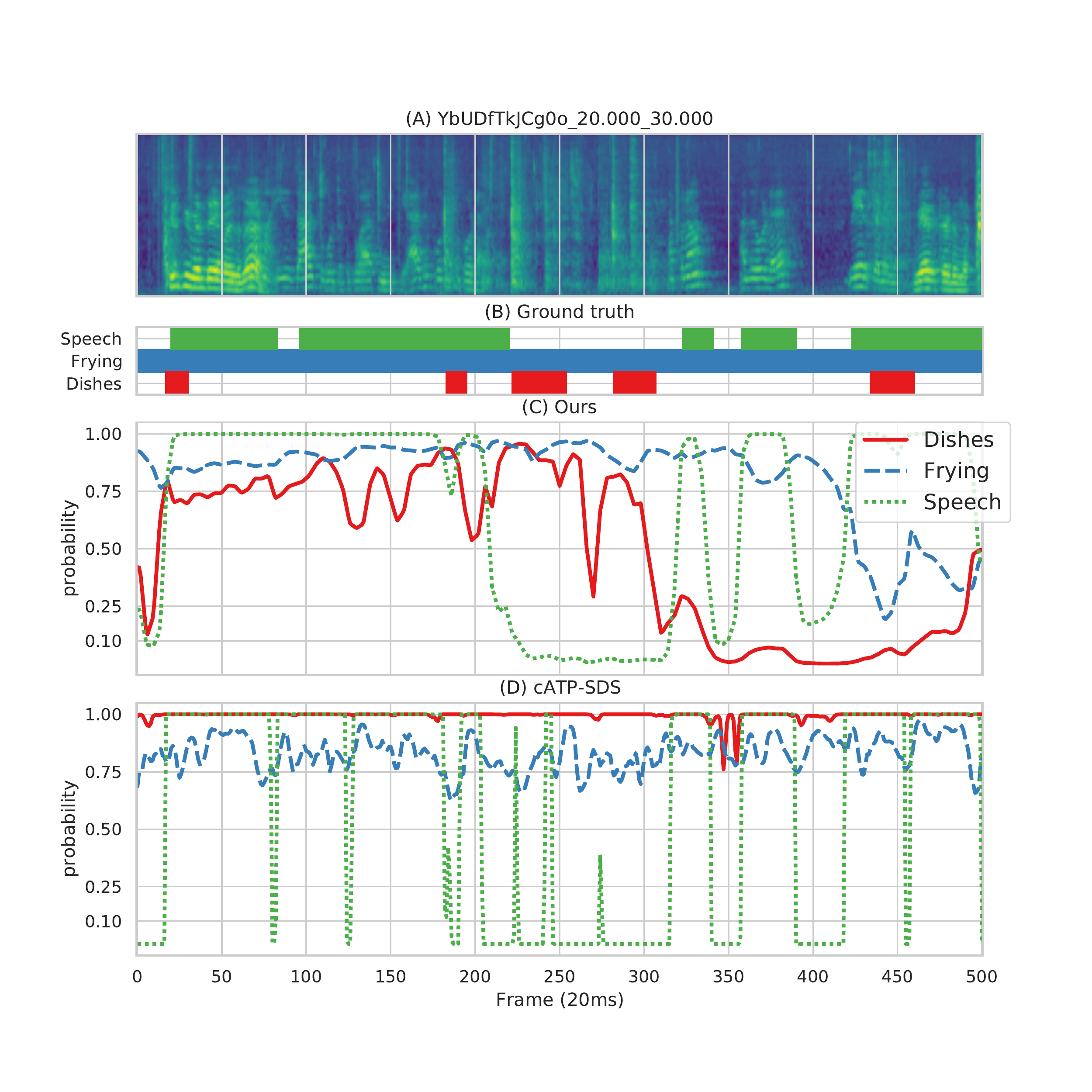}
    \caption{(A) A sample clip comparison on the (B) DCASE2018 evaluation dataset between (C) CDur and (D) cATP-SDS. Best viewed in color.}%
    \label{fig:quantative_dcase18}
\end{figure}

\begin{figure*}[htpb]
    \centering
    \includegraphics[width=0.98\textwidth]{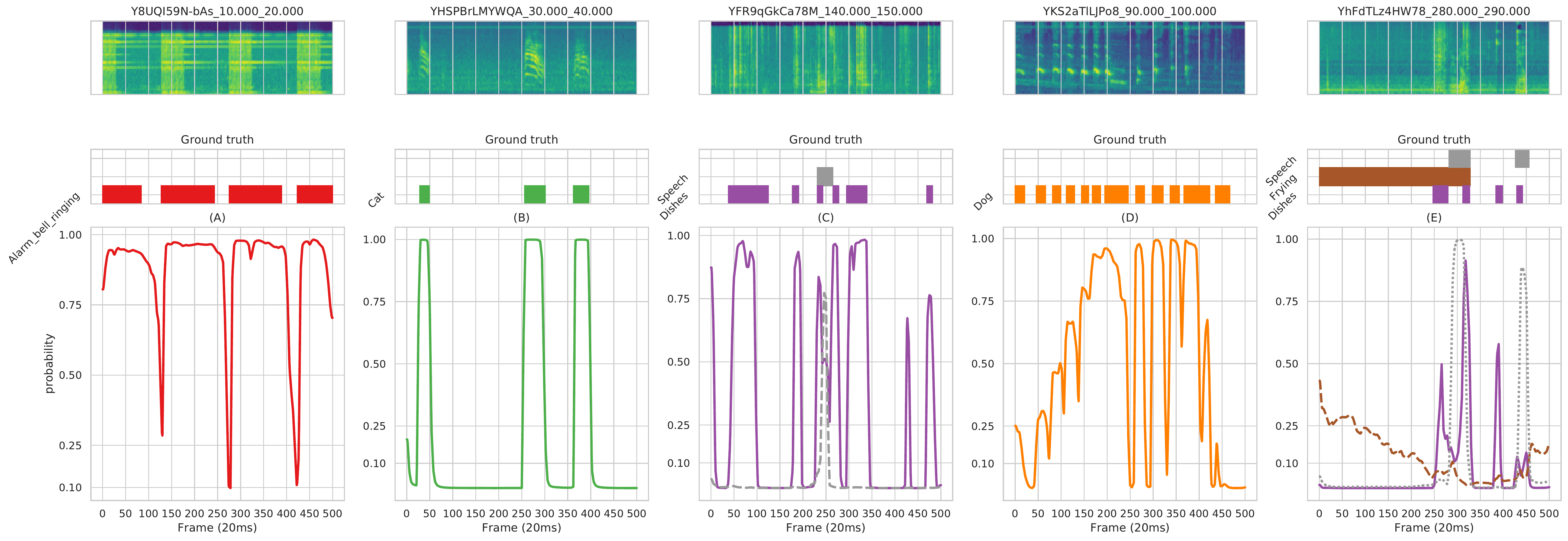}
    \caption{Predictions for five samples for each of the shortest duration events (``dishes'', ``cat'',``dog'',``speech'',``alarm bell ringing'') in DCASE2018. Best viewed in color.}%
    \label{fig:cdur_samples}
\end{figure*}

\begin{figure}[htpb]
    \centering
    \includegraphics[width=0.95\linewidth,trim={10 20 40 50},clip]{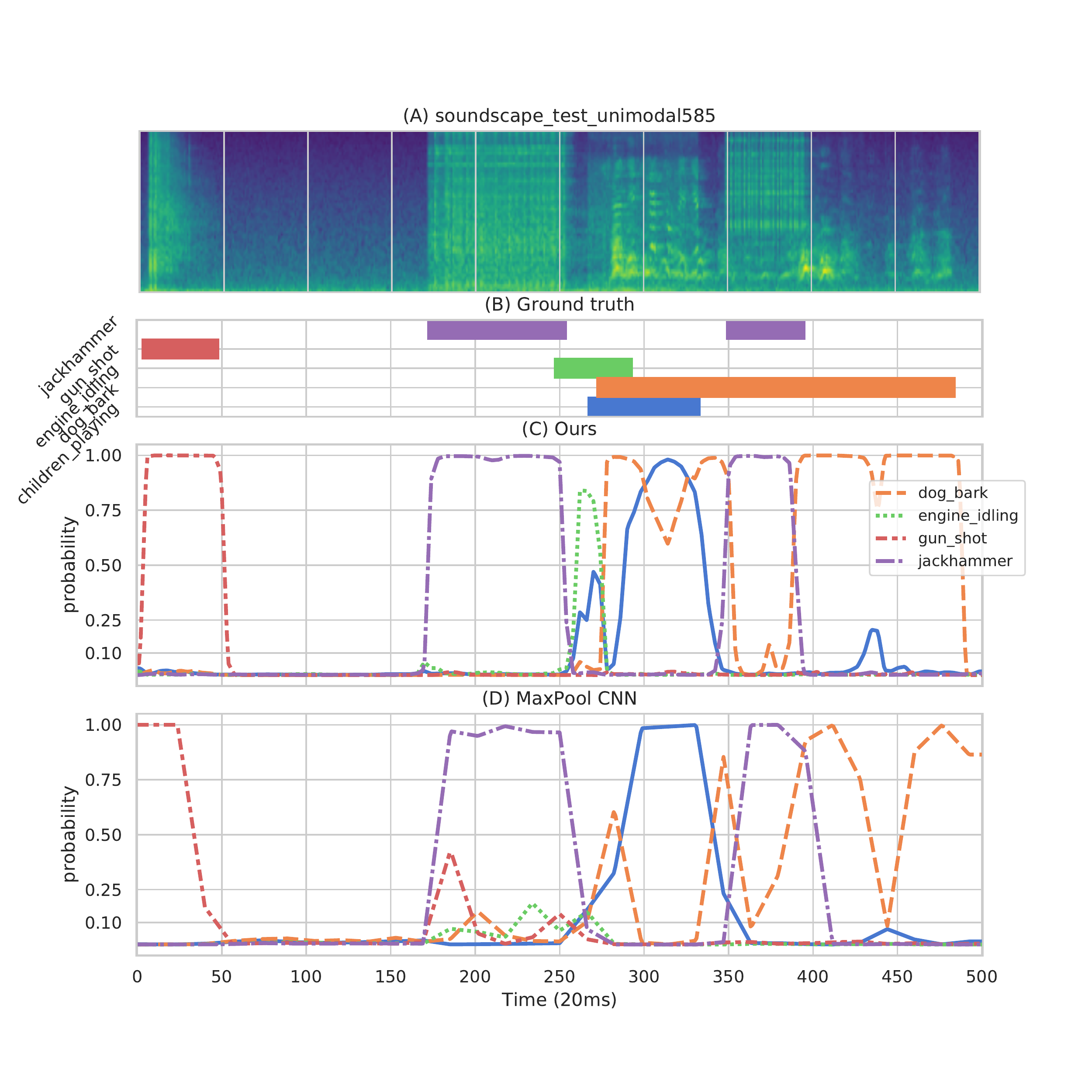}
    \caption{(A) A sample clip comparison on the (B) URBAN-SED evaluation dataset between (C) CDur and (D) MaxPool CNN. Best viewed in color.}%
    \label{fig:quantative_urbansed}
\end{figure}

Since our metrics are threshold-dependent, meaning that only hard labels were evaluated, we provide a quantitative analysis to strengthen the point of our models' duration robustness.
Two distinct clips containing at least three events from DCASE 2018 and URBAN-SED datasets are randomly sampled.
Regarding DCASE2018, our sampled clip (bUDfTkJCg0o) contains three different events (speech [green], frying [blue], dishes [red]).
We compare the probability outputs $y_t(e)$ of CDur (Event-F1 39.42) against cATP-SDS (Event-F1 34.07).
The comparison can be seen in \Cref{fig:quantative_dcase18}.
At first glance, it can be seen that both methods are incapable of producing perfect results.
Therefore we focus on the errors made by each individual approach.
In particular, for the speech event (green), one can notice that cATP-SDS predictions exhibit a peaking behavior with no apparent notion of the speech duration.
Therefore, cATP-SDS requires median filtering to remove false-positives and connect (or remove) its disjoint predictions.
In contrast, CDur seems to predict onset and offsets accurately without the need for post-processing.
This behavior is in accord with our previous observation in \Cref{tab:ablation}.
Additionally, the ``dishes'' event (a sharp cling sound of a fork hitting porcelain) is hard to estimate for both models.
However, cATP-SDS predicts dishes as an always present background noise, meaning that it failed to learn the characteristics of the short and sharp ``dishes'' sound.
Lastly, we also provide five distinct samples for the shortest duration events within DCASE2018 (``dishes'', ``cat'',``dog'',``speech'',``alarm bell ringing'') in \Cref{fig:cdur_samples}.
These samples further demonstrate that CDur is capable of detecting and accurately predicting short and sporadic events, such as ``alarm bell ringing'', ``cat'', and ``dishes''.
Our second sampled clip (soundscape\_test\_unimodal585) is from the URBAN-SED dataset (see \Cref{fig:quantative_urbansed}).
We compare CDur to a reimplementation of~\cite{McFee2018}, using our best performing Event-F1 MaxPool CNN model (\Cref{tab:urban_sed_results}).
Note that our reimplementation uses our LMS features with 50 Hz (20 ms/frame) frame rate, whereas the original work used 43 Hz (23 ms/frame).
At first glance, one can observe that this dataset is indeed artificial since the spectrogram appears to contain only the target events, without any other natural background noises.
As this sample shows, the MaxPool CNN model is indeed capable of sound localization, specifically for events with a duration of $\approx$ 1.5 s, such as ``jackhammer'', ``gun shot'' and ``children playing''.
However, it seems to struggle with longer events, such as ``dog bark'', at which it exhibits a peaking behavior, chunking the event into small pieces (see orange line).
On the contrary, CDur can predict and localize both short and long events for this sample.
Specifically, MaxPool CNN was unable to notice the short ``engine idling'' event (around 800 ms), yet CDur predicted its presence.
We believe that this is due to the low time-resolution of the MaxPool CNN model (320 ms/frame, 3.125 Hz), which could skip over the presence of short events. 

%\paragraph*{Discussion}

\section{Conclusion}
\label{sec:conclusion}

This work proposed CDur, a duration robust sound event detection CRNN model.
CDur aims to be as flexible as possible in order to be applied across different datasets and scenarios.
Further, we propose a new post-processing method called triple thresholding, which not only considers frame-level outputs but also utilizes the clip-level probabilities.
Triple thresholding can be seen to improve Event-F1 performance regarding Event-F1 performance on the DCASE2018 and URBAN-SED datasets.

CDur is then compared to other approaches in terms of Segment, Event, and Tagging performance.
Experiments conducted on the DCASE2017,18 and URBAN-SED datasets imply promising performance.
The DCASE2018 results show that CDur can outperform previous SOTA models in terms of Tagging- (69.11), Event- (39.42), and Seg-F1 (63.53).
Besides, on the URBAN-SED dataset CDur outperforms supervised methods in terms of Seg- (64.75) and Tagging-F1 (77.13).
A series of ablation experiments reveal the models' inherent capability to correctly localize sounds with effective onset and offset estimation. 
In our future work, we would like to investigate the new polyphonic scene detection score (psds)~\cite{Bilen2019} as a metric, which seems to provide promising insights towards a model's duration robustness and overall performance given a specific scenario.

%Therefore, CDur can be further modfied if necessary

% if have a single appendix:
%\appendix[Proof of the Zonklar Equations]
% or
%\appendix  % for no appendix heading
% do not use \section anymore after \appendix, only \section*
% is possibly needed

% use appendices with more than one appendix
% then use \section to start each appendix
% you must declare a \section before using any
% \subsection or using \label (\appendices by itself
% starts a section numbered zero.)
%

%\appendices
%\section{Proof of the First Zonklar Equation}
%Appendix one text goes here.

%% you can choose not to have a title for an appendix
%% if you want by leaving the argument blank
%\section{}
%Appendix two text goes here.

% use section* for acknowledgment
\section*{Acknowledgment}

This work has been supported by National Natural Science Foundation of China (No.61901265) and Shanghai Pujiang Program (No.19PJ1406300), and Startup Fund for Youngman Research at SJTU (No.19X100040009). Experiments have been carried out on the PI supercomputer at Shanghai Jiao Tong University.

% Can use something like this to put references on a page
% by themselves when using endfloat and the captionsoff option.
\ifCLASSOPTIONcaptionsoff
  \newpage
\fi

% trigger a \newpage just before the given reference
% number - used to balance the columns on the last page
% adjust value as needed - may need to be readjusted if
% the document is modified later
%\IEEEtriggeratref{8}
% The "triggered" command can be changed if desired:
%\IEEEtriggercmd{\enlargethispage{-5in}}

% references section

% can use a bibliography generated by BibTeX as a .bbl file
% BibTeX documentation can be easily obtained at:
% http://mirror.ctan.org/biblio/bibtex/contrib/doc/
% The IEEEtran BibTeX style support page is at:
% http://www.michaelshell.org/tex/ieeetran/bibtex/
\bibliographystyle{IEEEtran}
% \bibliographystyle{alphadin}
% argument is your BibTeX string definitions and bibliography database(s)
\bibliography{refs}
\end{document}